\documentclass[sigconf]{acmart}
\usepackage[utf8]{inputenc}
\renewcommand\footnotetextcopyrightpermission[1]{} 
\usepackage{graphicx}
\usepackage{balance}  
\usepackage{xcolor}

\usepackage{ifthen}
\usepackage{hyperref}
\usepackage{stmaryrd}
\usepackage{amsmath}
\usepackage{colortbl}
\usepackage{balance}
\usepackage{marginnote}
\usepackage{multicol}
\usepackage{listings}
\usepackage[linesnumbered]{algorithm2e}
\usepackage{algpseudocode}
\usepackage{mathtools}
\usepackage{bm}
\usepackage{caption}
\usepackage{enumerate}
\usepackage{lipsum}
\usepackage{verbatim}

\usepackage{tikz}
\usepackage{tkz-graph}
\usepackage{pgfplots}
\usepackage{subfig}
\usetikzlibrary{arrows, automata,petri,backgrounds}
\usetikzlibrary{positioning,shapes,shadows,arrows,decorations.markings,snakes}
\usetikzlibrary{decorations.pathmorphing}

\newcommand{\eat}[1]{}

\newcommand{\edgesin}[2]{N_{\rightarrow {#1}}^{#2}}
\newcommand{\edgesout}[2]{N_{{#1} \rightarrow}^{#2}}

\newtheorem{property}{Property}

\newcommand\mysubsection[1]{\vspace{-0mm}\subsection{#1}\vspace{-0mm}}
\newcommand\va{\vspace{-0mm}}

\newcommand\grow{\textsc{Grow}}
\newcommand\mergecur{\textsc{Merge}}

\newcommand{\extVersion}{false}
\newcommand{\printIfExtVersion}[2]
{
        \ifthenelse{\equal{\extVersion}{true}}{#1}{}
        \ifthenelse{\equal{\extVersion}{false}}{#2}{}
}

\copyrightyear{2020}
\acmYear{2020}
\acmConference[BDA 2020]{BDA 2020}{October 2020}{Paris, France}

\begin{document}
\title{Graph-based keyword search in heterogeneous data sources}
%
%

\author{Angelos~Christos~Anadiotis$^1$, Mhd~Yamen~Haddad$^2$,
Ioana~Manolescu$^2$}

\affiliation{
$^1$Ecole Polytechnique and Institut Polytechnique de Paris,
$^2$Inria and Institut Polytechnique de Paris,
}

\affiliation{$\{$name.surname$\}$ @ $^1$polytechnique.edu, $^2$inria.fr}
\renewcommand{\shortauthors}{A. C. Anadiotis, M.~Y.~Haddad and I.~Manolescu}

%

\begin{abstract}
Data journalism is the field of investigative journalism which focuses on digital data by treating them as first-class citizens.
Following the trends in human activity, which leaves strong digital traces, data journalism becomes increasingly important.
However, as the number and the diversity of data sources increase, heterogeneous data models with different structure, or even no structure at all, need to be considered in query answering.

Inspired by our collaboration with Le Monde, a leading French newspaper, we designed a novel query algorithm for exploiting such heterogeneous corpora through keyword search.
We model our underlying data as graphs and, given a set of search terms, our algorithm finds links between them within and across the heterogeneous datasets included in the graph.
We draw inspiration from prior work on keyword search in structured
and unstructured data, which we extend with the data heterogeneity dimension, which makes the keyword search problem computationally harder.
We implement our algorithm and we evaluate its performance using synthetic and real-world datasets.
\end{abstract}

\maketitle              

\section{Introduction}

Data analysis is increasingly important for several organizations today, as it creates value by drawing meaningful insights from the data.
As we are moving towards large data lakes installations where huge amounts of data are stored, the opportunities for important discoveries are growing; unfortunately, on par with the useless information.
Moreover, the data to be processed is often stored in different formats, ranging from fully and semi-structured, to completely unstructured, like free text.
Accordingly, the challenges in processing all this data that is available today, reside in both expressing and answering queries.

Research in heterogeneous data processing has proposed several approaches in addressing the above challenges.
On the one side, massively parallel processing systems like Spark~\cite{10.1145/2934664}, Hive~\cite{10.14778/1687553.1687609} and Pig~\cite{10.1145/1376616.1376726} provide connectors for heterogeneous data sources and allow the execution of data analysis tasks on top of them, using either a platform-specific API or a query language like SQL.
Polystore-based approaches~\cite{DBLP:conf/cidr/BugiottiBDIM15,10.1145/2463676.2463709,10.1145/2814710.2814713}  focus more on the data model and the query planning and optimization on top of heterogeneous data stores.
Finally, the so-called just-in-time (JIT) data virtualization approach generates the query engine at runtime based on the data format~\cite{DBLP:conf/cidr/KarpathiotakisA15, 10.14778/2994509.2994516}.
All these works consider that users, typically data scientists, already know what they are looking for, and they express it either using a powerful query language or a rich API.

However, today the data analysis paradigm has shifted and a central point is to find parts of the data which feature interesting patterns.
The patterns may not be known at query time; instead, users may have to discover them through a process of trial and error.
A popular query paradigm  in such a context is {\em keyword search}. A staple of Information Retrieval in data with little or no structure, keyword search has been applied also on relational, XML or graph data, when users are unsure of the structure and would like the system to identify possible connections. In this work,
we model a set of heterogeneous data sources as a graph, and focus on \textbf{answering  queries asking for connections among the nodes of the graph which are of interest to the users}.
This work is inspired from our collaboration with  Les Décodeurs, Le Monde's fact-checking team\footnote{http://www.lemonde.fr/les-decodeurs/}, within the ContentCheck collaborative research project\footnote{https://team.inria.fr/cedar/contentcheck/}.
Our study is novel with respect to the state of the art (Section~\ref{sec:related}) as we are the first to consider that an answer may span over multiple datasets of different data models, with very different or even absent internal structure, e.g., text data.
For instance, a national company registry is typically relational, contracts or political speeches are text, social media content typically comes as JSON documents, and open data is often encoded in RDF graphs.

\noindent\textbf{Integrated graph preserving all original nodes} In the data journalism context mentioned above, it is important to be able to {\em show where each piece of information in an answer came from}, and {\em how the connections were created}. This is a form of provenance, and can also be seen as result explanation. Therefore, the queried graph  needs to preserve the identity of each node from the original sources. At the same time, to enable interesting connections, we: ($i$)~extract several kinds of meaningful entities from all the data sources of all kinds; ($ii$)~interconnect data sources that comprise the same entity, or very similar ones, through so-called {\em sameAs}. Both extraction and similarity produce results with some {\em confidence}, a value between $0$ and $1$, thus, some edges in our graph have can be seen as uncertain (but quite likely).

\noindent\textbf{No help from a score function} An important dimension of keyword search problems is  {\em scoring}, i.e., how do we evaluate the interestingness of a given connection (or query result). This is important for two reasons. First, in many scenarios, the number of results is extremely large, users can only look at a small number of results, say $k$. Second, some answer score measures  have properties that can help limit the search, by allowing to determine that some of the answers not explored yet would not make it into the top $k$. Unfortunately, while desirable from an algorithmic perspective (since they simplify the problem), such assumptions on the score are not always realistic from a user perspective, as we learned by exchanging with journalists; we detail this in Section~\ref{sec:score}.

\noindent\textbf{Bidirectional search}  All edges in our graph are {\em directed}, e.g., from the subject to the object in an RDF graph, from the parent to the child in a hierarchical document etc., and, in keeping with our goal of integral source preservation, we store the edge direction in the graph. However, we allow answer trees to traverse edges in any direction, since heterogeneous data sources may model the same information either, say, of the form Alice $\xrightarrow{\text{wrote}}$ Paper$_1$ or Paper$_1$ $\xrightarrow{\text{hasAuthor}}$ Alice; since users are unfamiliar with the data, they should not be penalized for not having ``guessed'' correctly the edge directions. This is in contrast with many prior works (see Section~\ref{sec:related}) which define answers as a tree where from the root, a node matching each keyword is reached by traversing edges in their original direction only. For instance, assume the graph comprises $a_1\xrightarrow{\text{wrote}}p_1$ and $a_2\xrightarrow{\text{wrote}}p_1$. With a restricted notion of answers, the query $\{a_1\; a_2\}$ has no answer; in contrast, in our approach, the answer connecting them through $p_1$ is easily found. Bidirectional search gives a functional advantage, but makes the search more challenging: in a graph of $|E|$ edges, the search space is multiplied by $2^{|E|}$.

The contributions made in this work are as follows:
\begin{itemize}
\item We formalize the problem of bidirectional keyword search on graphs as described above, built from a combination of data sources.
\item With respect to scoring, we introduce a general score function that can be extended and customized to reflect all interesting properties of a given answer. We show that this generality, together with the possibility of confidence lower than $1.0$ on some edges, does not enable search to take advantage of simplifying assumptions made in prior work.
\item We propose a complete (if exhaustive) algorithm for solving the keyword search problem in this context, as well as some original pruning criteria arising specifically in the context of our graphs. Given the usually huge search space size, a practical use of this algorithm is to run it until a time-out and retain the best answers found. 
\item We have implemented our algorithm and present a set of experiments validating its practical interest.
\end{itemize}


A previous version of our system  had been demonstrated in~\cite{Chanial2018}. Since then, we have completely re-engineered the graph construction (this is described in the companion paper~\cite{construction-paper}), deepened our analysis of the query problem, and proposed a new algorithm, described in the present work; this also differs from (and improves over) our previous technical report~\cite{cordeiro:hal-02559688}.

\section{Outline and problem statement}
\label{sec:pb-statement}

In this section, we formalize our keyword search problem over a  graph that we build by integrating data from various datasets, organized in different data models.

\subsection{Integrated graph}
\label{sec:graph}
We consider a set $\mathcal M$ of \emph{data models}: relational (including SQL databases, CSV files etc.),  
RDF, 
JSON, 
HTML, XML, 
and text. A dataset $D$ is an instance of one of these data models\footnote{Our graph can also integrate other kinds of files, in particular PDF documents and spreadsheet files, by converting them to one or several instances of the above data models; as this is orthogonal wrt this paper, we delegate those details to~\cite{construction-paper}.}.

\begin{figure*}[t!]
\centering
\includegraphics[width=.9\textwidth]{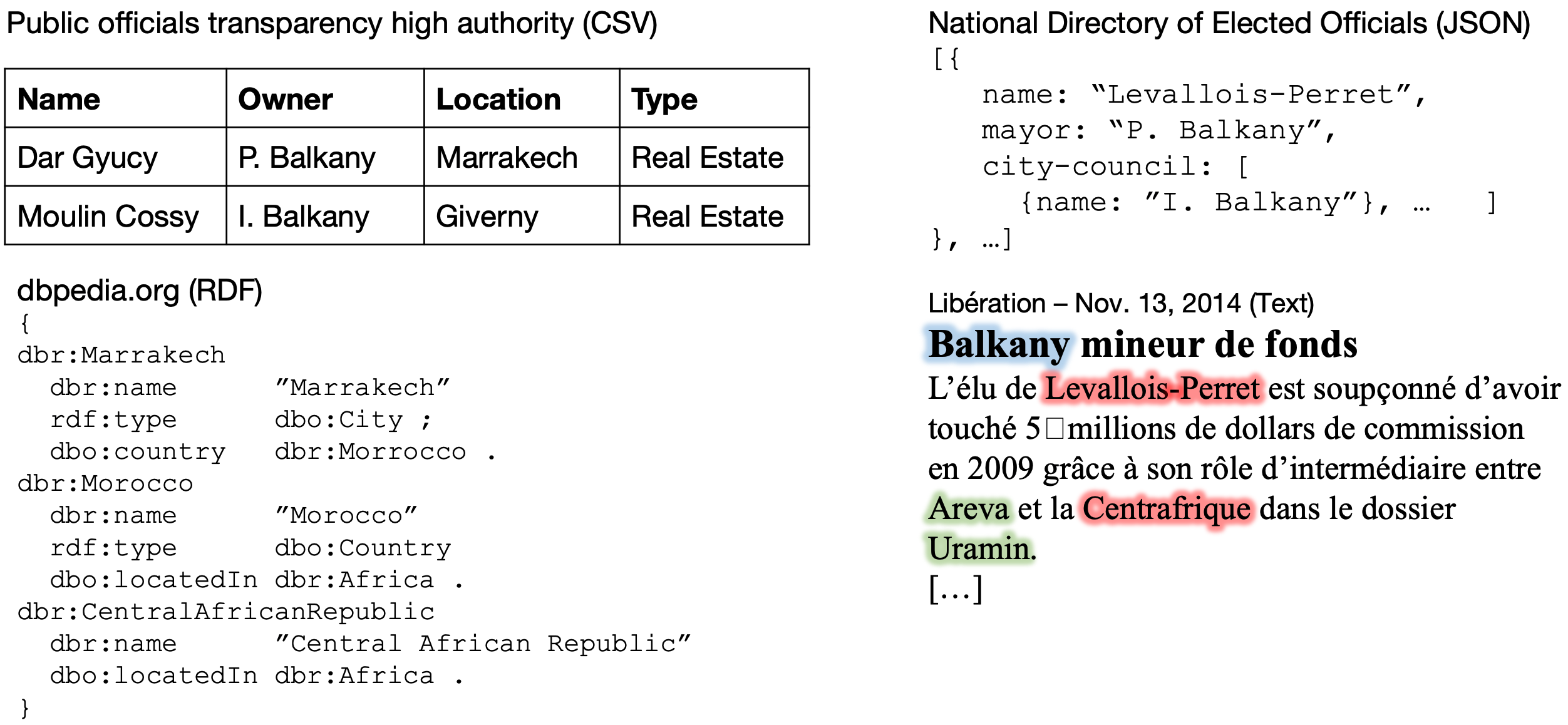}
\vspace{-3mm}
\caption{Sample dataset collection ${\mathcal D}$.\label{fig:example-ds}}
\end{figure*}

From a set $\mathcal{D}=\{D_1,D_2,\ldots, D_n\}$ of datasets, we create an \textbf{integrated graph}  $G=(N,E)$, where $N$ is the set of nodes and $E$ the set of edges. For instance, consider the dataset collection shown in Figure~\ref{fig:example-ds}. Starting from the top left, in clockwise order, it shows: a table with assets of public officials, a JSON  listing of France elected officials, an article from the newspaper Libération with entities highlighted, and a subset of the DBPedia RDF knowledge base.

\begin{figure*}[t]
\centering
\includegraphics[width=0.8\textwidth]{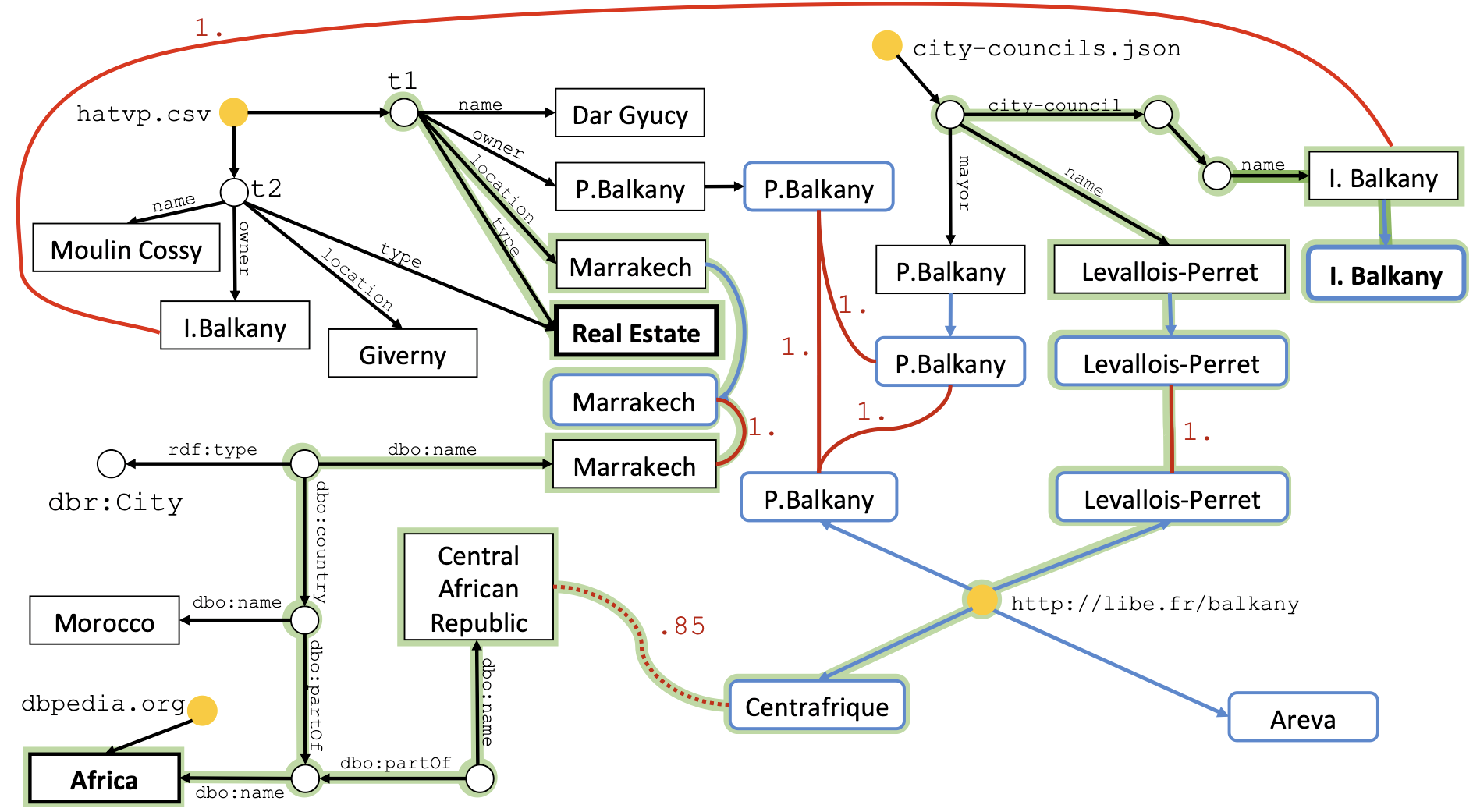}
\vspace{-2.5mm}
\caption{Integrated graph corresponding to the datasets of Figure~\ref{fig:example-ds}.\label{fig:example-graph}}
\end{figure*}

Figure~\ref{fig:example-graph} shows the graph produced from the datasets in Figure~\ref{fig:example-ds}.
There are several observations to be made on this graph:

($i$) The graph comprises four {\em dataset nodes} (the ones filled with yellow), one for each data source.

($ii$)~ {\em All the internal structure present in the input datasets is preserved} in the graph: each RDF node became a node in the integrated graph, and each triple became an edge. A node is created for each map, array, and value in the JSON document. A node is created from each tuple, and from each attribute in the relational databases. Finally, a single node is created from the whole text document, which has no internal structure. When a text consists of more than one phrase, we segment it as a \emph{sequence of phrases}, each of which is a node (child of the dataset node) to avoid overly large nodes that are hard to interpret by users.

($iii$)~{\em Entity nodes} (rounded-corners blue boxes) are extracted using Information Extraction (IE) techniques. Thus, in the example, nodes labeled ``P. Balkany'', ``I. Balkany'' are recognized as People, ``Levallois-Perret'' and ``Centrafrique'' are recognized as Locations, while ``Areva'' is an Organization. An extracted entity is added to the graph as a child of the node (leaf in an XML, HTML, JSON or text document; attribute value from a relational dataset; or RDF literal)  from which it has been extracted.

($iv$)~{\em Equivalence edges} (solid red  edges in Figure~\ref{fig:example-ds}) connect nodes found in different datasets which are considered to refer to the same real-world entity. For instance, the three occurrences of ``P.~Balkany'' are pairwise connected by edges with a {\em confidence} of $1.0$. The confidence of the edges derived directly from the datasets, as explained above, is $1.0$; we do not show it in the figure to avoid clutter.
We say nodes connected by equivalence edges are {\em equivalent}.

($v$)~{\em Similarity edges} (dotted, curved red edge between ``Central African Republic'' and ``Centrafrique'' in Figure~\ref{fig:example-ds}) connect nodes which are considered strongly similar but not equivalent. In our example, the two nodes have a similarity of $0.85$, which is attached to the edge as confidence.

For efficiency, when $k$ nodes are equivalent, we do not consider all the $\frac{k(k-1)}{2}$ edges; instead, one of the nodes (the first to be added to the graph - any other choice could be made) is designated the {\em representative} of all of them, and we store associated with each node, the ID of its representative.

The purpose of  the equivalence and similarity edges is to \textbf{interconnect nodes within and across the datasets}; entity extraction prepares the ground for the same, since it creates nodes that may co-occur across data sources, e.g., entities mentioned in separate texts, such as ``P. Balkany'' in the figure. This increases the value and usefulness of the graph, since it allows to find connections  which cannot be established based on any dataset taken separately.  For instance, consider the question: {\em \guillemotleft What connections exist between ``I. Balkany'', ``Africa'', and ``real estate''?\guillemotright} This can be asked as a \textbf{three-keyword query} \{``I. Balkany'', ``Africa", ``Estate''\}, for which an \textbf{answer} (a tree composed of graph edges) is shown as a light green highlight in Figure~\ref{fig:example-ds}; the three nodes matching the respective keywords are shown in bold. This answer interconnects all four data sources.

We formalize this keyword search query problem below.

\subsection{Search problem}
\label{sec:search-problem}

Given our graph $G=(N, E)$, we denote by $\mathcal{L}$ the set of all the labels of $G$ nodes, plus the special constant $\epsilon$ denoting the empty label. We denote by $\lambda(\cdot)$ a function assigning to each node and edge a label, which may be empty. As illustrated in Figure~\ref{fig:example-graph},  internal nodes, which correspond, e.g., to a relational tuple, or to a JSON map or array, have an empty label.


Let $W$ be the set of {\em keywords}, obtained by stemming the label set $\mathcal{L}$; a {\em search query} is a set of keywords $Q=\{w_1,...,w_m\}$, where $w_i\in W$.
We define an \textbf{answer tree} (AT, in short) as a set $t$ of  $G$ edges which ($i$)~together, form a tree (each node is reachable from any other through exactly one path), ($ii$)~for each $w_i$, contain at least one node whose label matches $w_i$. Here, the edges are \textbf{considered undirected}, that is: $n_1\xrightarrow{a}n_2\xleftarrow{b}n_3\xrightarrow{c}n_4$ is a sample AT,  such that for all $w_i \in Q$, there is a node  $n_i\in t$ such that $w_i \in \lambda(n_i)$.


We treat the edges of $G$ as undirected when defining the AT in order to allow more query results, on a graph built out of heterogeneous content whose structure is not well-known to users.  For instance, consider a query consisting of the keywords $k_1,k_4$ such that $k_1\in \lambda(n_1)$ and $k_4\in \lambda(n_4)$ on the four-nodes sample AT introduced above. If our ATs were restricted to the original direction of $G$ edges, the query would have no answer; ignoring the edge directions, it has one.
One could easily extend the definition and the whole discussion in order to allow matches to also occur on edges (just enlarge $\mathcal{L}$ to also include the stemmed edge labels).

Further, we are interested in  \textbf{minimal} answer trees, that is:

\begin{enumerate}
\item Removing an edge from the tree should make it lack one or more of the query keywords $w_i$.
\item \label{item:at-equiv} If a query keyword $w_i$ matches the label of more than one nodes in the answer tree, then all these matching nodes must be equivalent.
\end{enumerate}

Condition~(\ref{item:at-equiv}) is specific to the graph we consider, originating in {\em several data sources connected by equivalence or similarity edges}. In classical graph keyword search problems, each query keyword is matched {\em exactly once} in an answer (otherwise, the tree is considered non-minimal). In contrast, our answer trees \emph{may need to traverse equivalence edges}, and if $w_i$ is matched by one node connected by such an edge, it is also matched by the other. For instance, consider the three-keyword query ``Gyucy Balkany Levallois'' in Figure~\ref{fig:example-graph}:  the keyword Balkany is matched by the two nodes labeled ``P. Balkany'' which are part of the answer.

As a counter-example to condition (\ref{item:at-equiv}), consider the query ``Balkany Centrafrique'' in Figure~\ref{fig:example-graph}, assuming the keyword Centrafrique is also matched in the label ``Central African Republic''\footnote{This may be the case using a more advanced indexing system that includes some natural language understanding, term dictionaries etc.}.
Consider the tree that connects a ``P. Balkany'' node with ``Centrafrique'', and also traverses the edge between ``Centrafrique'' and ``Central African Republic'': this tree is not minimal, thus it is not an answer. The intuition for rejecting it is that  ``Centrafrique'' and ``Central African Republic'' may or may not be the same thing (we have a similarity, not an equivalence edge), therefore the query keyword ``Centrafrique'' is matched by two potentially different things in this answer, making it hard to interpret.


A direct consequence of minimality is that {\em in an answer, each and every leaf matches a query keyword}.

Several minimal answer trees may exist in $G$ for a given query. We consider available a {\em scoring function} which assigns a higher value to more interesting answer trees (see Section~\ref{sec:score}).
Thus, our problem can be stated as follows:

\begin{center}
\begin{tabular}{|p{0.9\columnwidth}|}
\hline
\textsc{Problem statement} Given the graph $G$  built out of the datasets  $\mathcal{D}$ and a query $Q$,  return the $k$ highest-score minimal answer trees.\\
\hline
\end{tabular}
\end{center}

An AT may potentially span over the whole graph, (also) because it can traverse $G$ edges in any direction;  this makes the problem challenging.

\noindent\textbf{Discussion: degraded answers.}
In some cases, a query may have no answer (as defined above) on a given graph, yet if one is willing to drop the second condition concerning  nodes matching the same query keyword, an answer tree could be found. For instance, consider a graph of the form $a_1\xrightarrow{l} b_1 \xrightarrow{m} b_2 \xrightarrow{n} c_1$, such that $b_1$ is \emph{not equivalent to} $b_2$, and the query $\{a, b, c\}$, such that the keyword $a$ matches the node $a_1$, $b$ matches $b_1$ and $b_2$ and $c$ matches $c_1$. Given our definition of answers above, this query has no answer, because $b$ matches the two nodes $b_1$ and $b_2$.

If we removed condition (2), we could accept such an answer, which we call {\em degraded}, since it is harder to interpret for users (lacking one clearly identified node for each keyword). One could then generalize our problem statement into: ($i$)~solve the problem stated above, and ($ii$)~only if there are no answers, find the top-$k$ degraded answers (if they exist). We do not pursue degraded answer search further in this paper, and focus instead on finding those defined above.


\subsection{Search space and complexity}
\label{sec:steiner}

\newcommand\pbma{$\diamond$}
\newcommand\pbmb{$\rhd$}
\newcommand\pbmc{$\lhd$}
\newcommand\pbmd{$\circ$}
\newcommand\pbme{$\Box$}

The problem that we study is related to the (Group) Steiner Tree Problem, which we recall below.

Given a graph $G$ with weights (costs) on edges, and a set of $m$ nodes $n_1,\ldots,n_m$, the \emph{Steiner Tree Problem (STP)} \cite{garey2011} consists of finding the tree in $G$ that connects all the nodes together, which minimizes the sum of the edge costs. We could answer our queries by solving one STP problem for each combination of nodes matching the keywords $w_1,\ldots,w_m$. However, there are several obstacles left: (\pbma)~STP is a known NP-hard problem in the size of $G$, denoted $|G|$; (\pbmb)~as we consider that each edge can be taken in the direct or reverse direction, this amounts to ``doubling'' every edge in $G$. Thus, our search space is  \textbf{$2^{|G|}$ larger than the one of the STP, or that considered in similar works}, discussed in Section~\ref{sec:related}. This is daunting even for small graphs of a few hundred edges;   (\pbmc)~we need the $k$ best trees, not just one; (\pbmd)~each keyword may match several nodes, not just one.

The closely related {\em Group STP} (GSTP, in short) \cite{garey2011} is: given $m$ {\em sets of nodes} from $G$, find the minimum-cost subtree connecting one node from each of these subtrees. GSTP does not raise the problem (\pbmd), but still has all the others.

In conclusion, the complexity of the problem we consider is extremely high. Therefore, solving it fully is unfeasible for large and/or high-connectivity graphs. Instead, our approach is:

\begin{itemize}
\item {\em Attempt to find all answers from the smallest} (fewest edges) {\em to the largest}. 
 Enumerating small trees first is both a practical decision (we use them to build larger ones) and fits the intuition that we shouldn't miss small answers that a human could have found manually.
However, as we will explain, we still ``opportunistically'' build some trees before exhausting the enumeration of smaller ones, whenever this is likely to lead faster to answers. The strategy for choosing to move towards bigger instead of smaller tress leaves rooms for optimizations on the search order.
\item {\em Stop at a given time-out or when $m$ answers have been found}, for some $m\geq k$;
\item {\em Return the $k$ top-scoring answers} found.
\end{itemize}

\noindent\textbf{Clarification about the cost.}
In our work, we do not consider the cost of a tree.
Instead, we employ a scoring function, which is orthogonal to the search algorithm and, thus, it does not require to make any assumptions about its monotonicity.



\section{Scoring Answer Trees}
\label{sec:score}

We now discuss how to evaluate the quality of an
answer. Section~\ref{sec:generic-score} introduces the general notion
of score on which we base our approach. Section~\ref{sec:specif}
describes one particular metric we attach to edges in order to
instantiate this score, finally Section~\ref{sec:concrete-score}
details the actual score function we used. 

\subsection{Generic score function}
\label{sec:generic-score}
We have configured our problem setting to allow {\em any scoring function}, which enables the use of different scoring schemes fitting the requirements of different users.
As a consequence, this approach allows us to study the interaction of the scoring function with different properties of the graph.
For instance, we are currently investigating the possibility to {\em learn} what makes an answer interesting for a user, so that we may return customized answers to each user.

Given an answer tree $t$ to a query $Q$, we consider a score function consisting of (at least) the following two components: 

\begin{itemize}
\item The {\em matching score} $ms(t)$, which reflects the quality of the answer
  tree, that is, how well its leaves match the query terms.
\item The {\em connection score} $cs(t)$, which reflects the quality of
  the tree connecting the edges. 
  Any formula can be used here, considering the number of edges, the confidence or any other property
attached to edges, or a query-independent property of the nodes, such as their PageRank or betweenness centrality score etc. 
\end{itemize}

The score of $t$ for $Q$, denoted $s(t)$, is computed as a 
combination of the two independent components $ms(t)$ and $cs(t)$. 
Popular combinations functions 
 (a weighted sums, or  product etc.) are monotonous in both
components, however, our framework does not require it. Finally, both $ms(t)$ and $cs(t)$ can be tuned based on a given user's preferences, to
personalize the score, or make them evolve in time through user feedback etc. 

\mysubsection{Edge specificity}\label{sec:specif}
We  now describe a metric on edges, which we used (through the
connection score $cs(t)$) to favor edges that are ``rare'' for both nodes they
connect. This metric was inspired by our experiments with real-world
data sources, and it helped return interesting answer trees in our
experience. 


For a given node $n$ and label $l$, let $\edgesin{n}{l}$ be the number of $l$-labeled edges entering $n$, and $\edgesout{n}{l}$ the number of $l$-labeled edges exiting $n$. 
\noindent The \textbf{specificity} of an edge $e=n_1\xrightarrow{l}n_2$ is defined as:

\begin{center}
\va\va
$s(e)=2/(\edgesout{n_1}{l} + \edgesin{n_2}{l})$.  
\va\va
\end{center}

$s(e)$ is $1.0$ for edges that are ``unique'' for both their source and their target, and decreases when the edge does not ``stand out'' among the edges of these two nodes.  
For instance,  the city council of Levallois-Perret comprises only one mayor (and one individual cannot be mayor of two cities in France, because he has to inhabit the city where he runs for office). Thus, the edge from the city council to P.~Balkany has a specificity of $2/(1.0+1.0)=1.0$. In contrast, there are 54 countries in Africa (we show only two), and each country is in exactly one continent;  thus,  
the specificity of the {\sf dbo:partOf } edges in the DBPedia fragment, going from the node named Morocco (or the one named Central African Republic) to the node named Africa is $2/(1+54)\simeq .036$.

\textbf{Specificity computation.}
When  registering the first dataset $D_1$, computing the specificity of its edges is trivial. 
However, when registering subsequent datasets $D_2,D_3$ etc., if some
node, say  $n_2\in D_2$ is found to be equivalent to a node $n_1\in
D_1$,  
{\em all} the $D_1$ edges adjacent to $n_1$ {\em and} the $D_2$ edges adjacent to $n_2$ should be reflected in the specificity of {\em each} of these edges. Thus,  in particular, the specificity of $D_1$ edges needs to be {\em recomputed} when a node in a source added after $D_1$ is equivalent to one of its nodes. 

A {\em na\"ive approach} would be: when the edges of $D_2$ are traversed (when we add this dataset to the graph), re-traverse the edges of $n_1$ in $D_1$ in order to (re)compute their specificity. However, that would be quite inefficient.

\begin{figure}[t!]
\begin{center}
\tikzstyle{arrow} = [thick,->,>=stealth]
\begin{tikzpicture}[node distance=1.3cm]
\node (zero) {$~$};
\node (n1) [right of=zero, xshift=-13mm, shape=circle, draw=black] {$n_1$};
\node (c1) [above of=n1, yshift=-8mm] {$es_1$}; 
\node (x) [right of=n1, xshift=8mm] {$x$}; 
\node (n2) [right of=n1, xshift=39mm] {$n_2$};
\draw [arrow, color=blue]  (x) -> (n1) node [midway, fill=white] {\textcolor{blue}{$l$}};
\node (dots1) [right of=n1, xshift=6mm, yshift=6mm] {$\ldots$};
\draw [arrow, color=blue]  (dots1) -> (n1) node [midway, fill=white] {\textcolor{blue}{$l$}};
\node (dots2) [right of=n1, xshift=6mm, yshift=-6mm] {$\ldots$};
\draw [arrow, color=blue]  (dots2) -> (n1) node [midway, fill=white] {\textcolor{blue}{$l$}};
\node (dots3) [left of=n2, , xshift=-6mm,yshift=6mm] {$\ldots$};
\draw [arrow, color=violet]  (dots3) -> (n2) node [midway, fill=white] {\textcolor{violet}{$l$}};
\node (dots4) [left of=n1, yshift=6mm] {$\ldots$};
\draw [arrow]  (n1) -> (dots4) node [midway, fill=white] {$b$};
\node (dots5) [left of=n1, yshift=-6mm] {$\ldots$};
\draw [arrow]  (dots5) -> (n1) node [midway, fill=white] {$c$};
\node (dots6) [right of=n2, yshift=6mm] {$\ldots$};
\draw [arrow]  (dots6) -> (n2) node [midway, fill=white] {$d$};
\node (dots7) [left of=n2, xshift=-6mm, yshift=-6mm] {$\ldots$};
\draw [arrow, color=violet]  (dots7) -> (n2) node [midway, fill=white] {\textcolor{violet}{$l$}};
\end{tikzpicture}
\caption{Illustration for specificity (re)computation. The specificity of the edge $x \xrightarrow{l}n_1$, $s(e)$ is initially computed out of the blue edges; when $n_2$ joins the equivalence set $es_1$, it is recomputed to also reflect the violet edges. \label{fig:edge-recomp}}
\end{center}
\vspace{-7mm}
\end{figure}
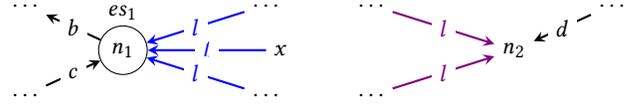

Instead, below, we describe an {\em efficient incremental algorithm} to compute specificity. We introduce two notations. 
For any edge $e$, we denote $N^e_{\rightarrow \bullet}$, respectively $N^e_{\circ \rightarrow}$,  the two  numbers out of which the specificity of $e$ has been {\em most recently} computed\footnote{This can be either during the first specificity computation of $e$, or during a recomputation,  as discussed below.}. Specifically,  $N^e_{\rightarrow \bullet}$ counts $l$-labeled edges incoming to the target of $e$, while $N^e_{\circ \rightarrow}$ counts $l$-labeled edges outgoing the source of $e$. In Figure~\ref{fig:edge-recomp}, if $e$ is the edge $x \xrightarrow{l}n_1$, then $N^e_{\rightarrow \bullet}=3$ (blue edges) and $N^e_{\circ \rightarrow}=1$, thus $s(e)=2/4=.5$.  

Let $n_1\in D_1$ be a node, $es_1$ be the set of all nodes equivalent to
$n_1$, and $n_2\in D_2$ be a node in a dataset we currently register,
and which has just been found to be equivalent to $n_1$, also. 

Further, let $l$ be a label of an edge incoming or outgoing  (any) node from $es_1$, and/or $n_2$.
We denote by $\edgesin{es_1}{l}$ the sum $\sum_{n\in es_1}(\edgesin{n}{l})$  and similarly by $\edgesout{es_1}{l}$ the sum $\sum_{n\in es_1}(\edgesout{n}{l})$; they are the numbers of $l$-labeled outgoing (resp., incoming) $l$-labeled edges of any node in $es_1$. 
When $n_2$ joins the equivalence set $es_1$ of $n_1$ (see Figure~\ref{fig:edge-recomp}): 

\va\va
\begin{enumerate}
\item \label{item:incremental-1}
If $\edgesin{es_1}{l}\neq 0$ and $\edgesin{n_2}{l}\neq 0$, the specificity of every $l$-labeled edge $e$  {\em incoming} either a node in $es_1$ or the node $n_2$ must be recomputed.\\
Let $e$ be such an {\em incoming} edge labeled $l$. When $n_2$ is added to the set $es_1$, the specificity of $e$  becomes 
%
$2/((N^e_{\rightarrow \bullet} + \edgesin{n_2}{l}) + N^e_{\circ \rightarrow})$, 
%
to reflect that $n_2$ brings more incoming $l$-labeled edges. This amounts to $2/(3+2+1)=.33$ in Figure~\ref{fig:edge-recomp}: the violet edges have joined the blue ones. Following this adjustment, the numbers out of which $e$'s specificity has been most recently computed are modified as follows: 
%
$N^e_{\rightarrow \bullet}$  becomes $N^e_{\rightarrow \bullet} + \edgesin{n_2}{l}$, thus $3+2=5$ in Figure~\ref{fig:edge-recomp}; 
$N^e_{\circ \rightarrow}$ remains unchanged.
\va
\item \label{item:incremental-2} 
If $\edgesin{es_1}{l}=0$ and $\edgesin{n_2}{l}\neq 0$, the specificity of every $l$-labeled edge $e$  {\em incoming} $n_2$ 
does not change when $n_2$ joins the equivalence set $es_1$. 
\item \label{item:incremental-3}
If $\edgesin{es_1}{l}\neq 0$ and $\edgesin{n_2}{l}=0$, the newly added node $n_2$ does not change the edges adjacent to the nodes of $es_1$, nor their specificity values.

\printIfExtVersion{
\noindent\textbf{4. }
Similarly to \textbf{1.} 
above, if $\edgesout{es_1}{l}\neq 0$ and $\edgesout{n_2}{l}\neq 0$, the specificity of every $l$-labeled edge $e$  {\em outgoing} either a node in $es_1$ or the node $n_2$ is recomputed, and it becomes 
%
$2/(N^e_{\rightarrow \bullet} + (N^e_{\circ \rightarrow}+\edgesout{n_2}{l}))$. 
%
The numbers out of which $e$'s specificity has been most recently computed become: $N^e_{\rightarrow \bullet}$  and  $(N^e_{\circ \rightarrow}+\edgesout{n_2}{l})$.
%
\noindent\textbf{5. }
Finally, if  $\edgesout{es_1}{l}=0$ and $\edgesout{n_2}{l}\neq 0$, the reasoning is similar to the one in \textbf{2. } 
above, while if  $\edgesout{es_1}{l}\neq 0$ and $\edgesout{n_2}{l}=0$, it is similar to \textbf{3.}; 
in both cases, we consider outgoing (instead of incoming) edges and edge count numbers.
}{
\end{enumerate}
\va
The last two cases, when $\edgesout{es_1}{l}\neq 0$ and $\edgesout{n_2}{l}\neq 0$, respectively, $\edgesout{es_1}{l}=0$ and $\edgesout{n_2}{l}\neq 0$, are handled in a similar manner. 
}

The above method only needs, for a given node $n_2$ newly added to the graph, and label $l$, the number of edges adjacent to $n_2$ in its dataset, and the \emph{number} of $l$ edges  adjacent to a node equivalent to $n_2$. Unlike the na\"ive specificity computation method,  it does not need to actually {\em traverse} these edges previously registered edges, making it more efficient.  

Concretely, for each edge $e\in E$, we store three attributes: $N^e_{\rightarrow \bullet}$, $N^e_{\circ \rightarrow}$  and $s$, the last-computed specificity, and we update $N^e_{\rightarrow \bullet}$, $N^e_{\circ \rightarrow}$ as explained above. 

\subsection{Concrete score function}
\label{sec:concrete-score}
In our experiments, we used the following score function.


For an answer $t$ to the query $Q$, we compute the matching score $ms(t)$  as the \emph{average}, over all query keywords $w_i$, of the similarity between the $t$ node matching $w_i$ and the keyword $w_i$ itself; we used the edit distance. 

We compute the connection score $cs(t)$ based on edge confidence, on one hand, and edge specificity on the other. We {\em multiply} the confidence values, since we consider that uncertainty (confidence $<1$) multiplies; and we also {\em multiply}  the specificities of all edges in $t$, to discourage many low-specificity edges. Specifically, our score is computed as: 



\va\va
\begin{center}
$score(t,Q)=\alpha \cdot ms(t,Q) + \beta \cdot \prod_{e\in  E} c(e) + (1 - \alpha - \beta) \cdot \prod_{e\in E} s(e)$
\end{center}
\va\va

\noindent where $\alpha$, $\beta$ are parameters of the system such that $0\leq \alpha, \beta <1$ and $\alpha +\beta\leq 1$.

\subsection{Orthogonality between the score and the algorithm}

Before we describe the search algorithm, we make a few more remarks on the connection between the score function and the search algorithm. 

We start by considering the classical Steiner Tree and Group Steiner Tree Problems (Section~\ref{sec:steiner}). These assume that the score is \textbf{monotonous}, that is: for any query $Q$ and all trees $T_1, T_2$ where $T_1$ is a subtree of $T_2$, the score of $T_2$ is lower than that of $T_1$. This is naturally satisfied if the score is the addition of edge weights.

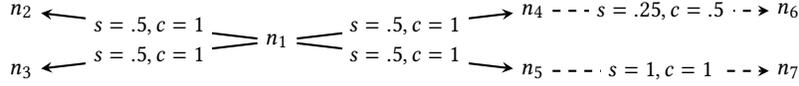
\begin{figure*}[t!]
\begin{center}
\tikzstyle{node} = [text centered, fill=white]
\tikzstyle{arrow} = [thick,->,>=stealth]
\begin{tikzpicture}[node distance=34mm and 34mm]
\node (zero) {~}; 
\node (n1) [node, right of=zero, xshift=-14mm] {$n_1$};
\node (n2) [node, left of=n1, yshift=4mm] {$n_2$};
\node (n3) [node, left of=n1, yshift=-4mm] {$n_3$};
\node (n4) [node, right of=n1, yshift=4mm] {$n_4$};
\node (n5) [node, right of=n1, yshift=-4mm] {$n_5$};
\draw [arrow] (n1) -> (n2) node [midway, fill=white] {$s=.5,c=1$};
\draw [arrow] (n1) -> (n3) node [midway, fill=white] {$s=.5,c=1$};
\draw [arrow] (n1) -> (n4) node [midway, fill=white] {$s=.5,c=1$};
\draw [arrow] (n1) -> (n5) node [midway, fill=white] {$s=.5,c=1$};
\node (n6) [node, right of=n4] {$n_6$};
\node (n7) [node, right of=n5] {$n_7$};
\draw [arrow, dashed] (n4) -> (n6) node [midway, fill=white] {$s=.25,c=.5$}; 
\draw [arrow, dashed] (n5) -> (n7) node [midway, fill=white] {$s=1,c=1$}; 
\end{tikzpicture}
\caption{Example (non-monotonicity of the tree score). $T$ is the four-edges tree rooted in $n_1$.}
\label{fig:nonmono}
\end{center}
\end{figure*}

However, in its general form  (Section~\ref{sec:generic-score}), and in particular our concrete score  (Section~\ref{sec:concrete-score}), is \textbf{not monotonous}, as illustrated  in Figure~\ref{fig:nonmono}, where on each edge, $c$ is the confidence and $s$ is the specificity. 
Let $T_1$ denote the four-edge tree rooted in $n_1$, $T_2$ be the five-edges tree consisting of $T_1$ plus the edge from $n_4$ to $n_6$, and $T_3$ be the five-edges tree consisting of $T_1$ plus the edge from $n_5$ to $n_7$. Assume $\alpha=\beta=\frac{1}{3}$ and that $T_1,T_2,T_3$ have the same matching score.  Then, the last two terms in their score are as follows:

\begin{itemize}
  \item  $score(T_1,Q)=\alpha \cdot ms(T_1,Q) + \beta \cdot (.5)^4 + \gamma$;  
  \item $score(T_2,Q) = \alpha \cdot ms(T_2,Q) + \beta (.5)^4\cdot 0.25 +\gamma \cdot .5$;
  \item $score(T_3,Q)= \alpha \cdot ms(T_3,Q) + \beta (.5)^4 + \gamma$
  \end{itemize}

For any non-zero $\alpha$, if $ms(T_3,Q)> ms(T_1,Q)$, then $score(T_3,Q)> score(T_1,Q)$, contradicting the monotonicity assumption. For what concerns $T_2$, the last two score components are lower than  $T_1$'s; whether  $T_2$'s score is higher than lower than that of $T_1$ depends on their matching score and on the chosen coefficients.
  
%
%



Another property sometimes assumed by score functions is the called  \textbf{optimal substructure}, that is: the best solution for a problem of size $p$ is part of the best solution for a problem of size $p+1$ that is an extension of $p$, for some problem size $p$. When this holds, the problem can be efficiently solved in a dynamic programming fashion. 
However, STP  does not enjoy this property: the best-score tree connecting two nodes $n_1,n_2$ is not necessarily part of the best-score tree that connects $n_1,n_2,n_3$ (and the same holds for GSTP). Some existing algorithms also assume a variant of the optimal substructure property (see Section~\ref{sec:related}). In contrast, our score function  does not ensure such favorable properties. This is why the search algorithm we describe next has to find as many answers as possible, as quickly as possible. 

\section{Answering keyword queries}
\label{sec:algo}

We now present our approach for computing query answers, based on the integrated graph.

\subsection{\grow\ and  \mergecur}
\label{sec:algorithm}

Our first algorithm uses some concepts from the prior literature~\cite{dpbf,blinks} while exploring many more trees. Specifically, it starts from the sets of nodes $N_1,\ldots,N_m$ where the nodes in $N_i$ all match the query keyword $w_i$; each node $n_{i,j}\in N_i$ forms a one-node partial tree. For instance, in Figure~\ref{fig:example-graph}, one-node trees are built from the nodes with boldface text, labeled ``Africa'', ``Real Estate'' and ``I. Balkany''.  Two transformations can be applied to form increasingly larger trees, working toward query answers:

\begin{itemize}
\item \grow($t,e$), where $t$ is a tree, $e$ is an edge \emph{adjacent to the root} of $t$, and $e$ does not close a loop with a node in $t$, creates a new tree $t'$ having all the edges of $t$ plus $e$; the root of the new tree is the other end of the edge $e$. For instance, starting from the node labeled ``Africa'', a \grow\ can add the edge labeled {\small \texttt{dbo:name}}.
\item \mergecur($t_1,t_2$), where $t_1,t_2$ are trees with the same root, whose other nodes are disjoint, and
matching disjoint sets of keywords, creates a tree $t''$ with the same root and with all edges from $t_1$ and $t_2$. 
Intuitively, \grow\ moves away from the keywords, to explore the graph; \mergecur\ fuses two trees into one that matches more keywords than both $t_1$ and $t_2$.
\end{itemize}

In a {\em single-dataset} context, \grow\ and \mergecur\  have the following properties.
($gm_1$)~\grow\ alone is \textbf{complete} (guaranteed to find all answers) for $k=1,2$ only; for higher $k$, \grow\ and \mergecur\, together are complete.
($gm_2$)~Using \mergecur\ steps helps find answers faster than using just \grow~\cite{blinks}: partial trees, each starting from a leaf that matches a keyword, are merged into an answer as soon as they have reached the same root.
($gm_3$)~An answer can be found through \textbf{multiple combinations of \grow\ and \mergecur}. For instance, consider a linear graph $n_1\rightarrow n_2 \rightarrow \ldots n_p$ and the two-keyword query $\{a_1, a_p\}$ where $a_i$ matches the label of $n_i$. The answer is obviously the full graph. It can be found: starting from $n_1$ and applying $p-1$ \grow\ steps; starting from $n_p$ and applying $p-1$ \grow\ steps; and in $p-2$ ways of the form \mergecur(\grow(\grow\ldots), \grow(\grow\ldots)), each merging in an intermediary node $n_2,\ldots,n_{p-1}$. These are all the same according to our definition of an answer (Section~\ref{sec:search-problem}), which does not distinguish a root in an answer tree; this follows users' need to know how things are connected, and for which the tree root is irrelevant.

\subsection{Adapting to multi-datasets graphs}
The changes we brought for our harder problem (bidirectional edges and multiple interconnected datasets) are as follows.

~\\
\noindent\textbf{1. Bidirectional growth.} We allow \grow\ to traverse an edge both going from the source to the target, and going from the target to the source. For instance, the {\sf \small type} edge from ``Real Estate'' to {\sf \small $<$tuple1$>$} is traversed target-to-source, whereas the {\sf \small location} edge from {\sf \small $<$tuple1$>$} to ``Real Estate'' is traversed source-to-target.

\vspace{3mm}
\noindent\textbf{2. Many-dataset answers.} As defined in a single-dataset scenario, \grow\ and \mergecur\ do not allow to connect multiple datasets. To make that possible, we need to enable one, another, or both to also traverse similarity and equivalence edges (shown in solid or dotted red lines in Figure~\ref{fig:example-graph}. We decide to simply extend \grow\ to allow it to traverse not just data edges, but also {\em similarity} edges between nodes of the same or different datasets.
We handle {\em equivalence} edges as follows: 


\vspace{1.5mm}
\noindent\textbf{2.a Na\"ive solution: \grow-to-equivalent.} The simplest idea is to allow \grow\ to also add an equivalence edge to the root of a tree. However, this can be very inefficient. Consider three equivalent nodes $m$, $m'$ and $m''$, e.g., the three ``P. Balkany'' nodes in Figure~\ref{fig:example-graph}: a \grow\ step could add one equivalence edge, the next \grow\ could add another on top of it etc. More generally, for a group of $p$ equivalent nodes, from a tree rooted in one of these nodes, $2^p$ trees  would be created just by \grow. In our French journalistic datasets, some entities, e.g. ``France'', are very frequent, leading to high $p$;  exploring $2^p$ subtrees every time we reach a ``France'' node is extremely expensive\footnote{Note that {\em similarity} edges do not raise the same problem, because in our graph we only have such edges if the similarity between two nodes is above a certain threshold $\tau$. Thus, if a node $n_1$ is at least $\tau$-similar to $n_2$, and $n_2$ is at least $\tau$-similar to $n_3$, $n_1$ may be at least $\tau$-similar to $n_3$, or not. This leads to much smaller groups of similar nodes, than the groups of equivalent nodes we encountered.}.

\newcommand{\gtr}{\textsc{Grow2Rep}}
\vspace{1.5mm}
\noindent\textbf{2.b \grow-to-representative}
To avoid this, we devise a third algorithmic step, called {\em \grow-to-representative} (\gtr), as follows. Let $t$ be a partial tree developed during the search, rooted in a node $n$, such that the  representative of $n$ (recall Section~\ref{sec:graph}) is a node $n_{rep}\neq n$.  \gtr\ creates a new tree by adding to $t$ the edge $n\xrightarrow{\equiv}n_{rep}$; this new tree is rooted in $n_{rep}$. If $n$ is part of a group of $p$ equivalent nodes,  only  {\em one} \gtr\ step is possible from $t$, to the unique representative of $n$; \gtr\ does not apply again on \gtr($t$), because the root of this tree is $n_{rep}$, which is its own representative.

Together, \grow, \gtr\ and \mergecur\ enable finding answers that span multiple data sources, as follows:
\begin{itemize}
\item \grow\ allows exploring data edges within a dataset, and similarity edges within or across datasets;
\item \gtr\ goes from a node to its representative when they differ; the representative may be in a different dataset;
\item \mergecur\ merges trees with a same root: when that root is the representative of a group of $p$ equivalent nodes, this allows connecting partial trees, including \gtr\ results, containing nodes from different datasets. Thus, \mergecur\ can build trees spanning multiple datasets.
\end{itemize}

One potential performance problem remains. Consider again $p$ equivalent nodes $n_1,\ldots,n_p$; assume without loss of generality that their representative is $n_1$. Assume that during the search, a tree $t_i$ is created rooted in each of these $p$ nodes. \gtr\ applies to all but the first of these trees, creating the trees $t_2', t_3', \ldots, t_p'$, all rooted in $n_1$. Now, \mergecur\ can merge any pair of them, and can then repeatedly apply to merge three, then four such trees etc., as they all have the same root $n_1$. The exponential explosion of \grow\ trees, avoided by introducing \gtr, is still present due to \mergecur!

We solve this problem as follows.
Observe that in an answer, a  {\em path of two or more equivalence edges} of the form $n_1\xrightarrow{\equiv}n_2\xrightarrow{\equiv}n_3$ such that {\em a node internal to the path}, e.g. $n_2$, {\em has no other adjacent edge}, even if allowed by our definition, is \emph{redundant}. Intuitively, such a node brings nothing to the answer, since its neighbors, e.g., $n_1$ and $n_3$, could have been connected directly by a single equivalence edge, thanks to the transitivity of equivalence. We call {\em non-redundant} an answer that does not feature any such path, and decide to \textbf{search for non-redundant answers} only.


The following properties hold on non-redundant answers:

\begin{property}
There exists a graph $G$ and a $k$-keyword query $Q$ such that a non-redundant answer contains $k-1$ adjacent equivalence edges
(edges that, together, form a single connected subtree).
\end{property}

\begin{figure}[h!]
\vspace{-2mm}
\centering
\includegraphics[width=.9\columnwidth]{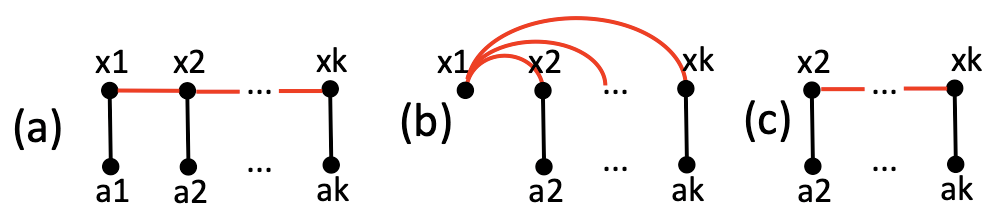}
\vspace{-2mm}
\caption{Sample answer trees for algorithm discussion.\label{fig:prop-example}}
\end{figure}

We prove this by exhibiting such an instance.  Let $G$ be a graph of $2k$ nodes shown in Figure~\ref{fig:prop-example} (a),  such that all the $x_i$ are equivalent, 
and consider the $k$-keyword query $Q=\{a_1,\ldots,a_k\}$ (each keyword matches exactly the respective $a_i$ node).  An answer needs to traverse all the $k$ edges from $a_i$ to $x_i$, and then connect the nodes $x_i,\ldots,x_k$; we need $k-1$ equivalence edges for this.

Next, we show:

\begin{property}\label{prop:2}
Let $t$ be a non-redundant answer to a query $Q$ of $k$ keywords. A group of adjacent equivalence edges contained in $t$  has at most $k-1$ edges.
\end{property}

We prove this by induction over $k$.
For $k=1$, each answer has $1$ node and $0$ edge (trivial case).

Now, consider this true for $k$ and let us prove it for $k+1$. Assume by contradiction that a non-redundant answer $t_Q$ to a query $Q$ of $k+1$ keywords comprises $k+1$ adjacent equivalence edges. Let $Q'$ be the query having only the first $k$ keywords of $Q$, and $t'$ be a subtree of $t$ that is a non-redundant answer to $Q'$:
\begin{itemize}
\item $t'$ exists, because $t$ connects all $Q$ keywords, thus also the $Q'$ keywords;
\item $t'$ is non-redundant, because its edges are also in the (non-redundant) $t$.
\end{itemize}

By the induction hypothesis, $t'$ has at most  $k-1$ adjacent equivalence edges. This means that there are {\em two adjacent equivalent edges} in $t\setminus t'$.  

\begin{enumerate}
\item If these edges, together, lead to two distinct leaves of $t$, then $t$ has {\em two} leaves  not in $t'$. This is not possible, because by definition of an answer, $t$ has $k+1$ leaves (each matching a keyword) and similarly $t'$ has $k$ leaves.
\item It follows, then, that the two edges lead to a single leaf of $t$, therefore the edges form a redundant path. This contradicts the non-redundancy of $t$, and
concludes our proof.
\end{enumerate}

Property~\ref{prop:2} gives us an important way to control the exponential development of trees due to $p$ equivalent nodes. \grow, \gtr\ and \mergecur, together, can generate trees with up to $k$ (instead of $k-1$) adjacent equivalence edges. 
This happens because \gtr\ may ``force'' the search to visit the representative of a set of $k$ equivalent nodes (see Figure~\ref{fig:prop-example}(b), assuming $x_1$ is the representative of all the equivalent $x_i$s, and the query $\{a_2,\ldots,a_k\}$). The resulting answer may  be  redundant, if the representative has no other adjacent edges in the answer other than equivalence edges. In such cases, 
in a \textbf{post-processing step}, we remove  from the answer the representative and its equivalence edges, then reconnect  the respective equivalent nodes using $k-1$ equivalence edges. This guarantees obtaining a non-redundant tree, such as the one in Figure~\ref{fig:prop-example}(c).

\subsection{The GAM algorithm}

\begin{figure*}[t!]
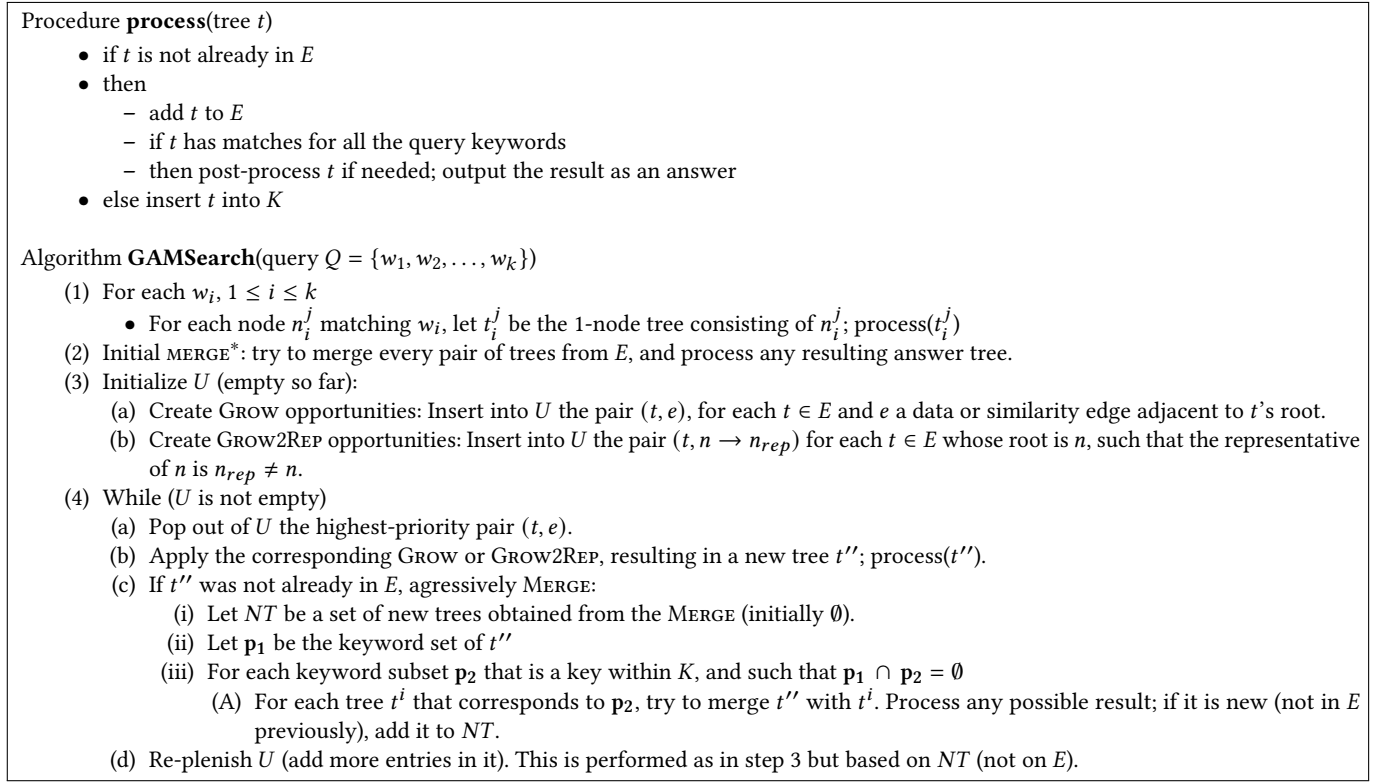

\fbox{
\begin{minipage}{\textwidth}
Procedure \textbf{process}(tree $t$)
\begin{itemize}
\item if $t$ is not already in $E$
 \item then
\begin{itemize}
\item add $t$ to $E$
\item if $t$ has matches for all the query keywords
\item then post-process $t$ if needed; output the result as an answer
\end{itemize}
\item else insert $t$ into $K$
\end{itemize}
~\\
Algorithm \textbf{GAMSearch}(query $Q=\{w_1,w_2,\ldots, w_k\}$)
\begin{enumerate}
\item For each $w_i$, $1\leq i \leq k$
\begin{itemize}
\item For each node $n_i^j$ matching $w_i$, let $t_i^j$ be the 1-node tree consisting of $n_i^j$; process($t_i^j$) \label{item:1node}
\end{itemize}
 \item Initial \textsc{merge}$^*$: try to merge every pair of trees from
   $E$, and process any resulting answer tree. \item Initialize $U$ (empty so far):  \label{item:init-Q}
\begin{enumerate}
\item Create \grow\ opportunities: Insert into $U$ the pair $(t,e)$, for each $t\in E$ and $e$ a data or similarity edge adjacent
   to $t$'s root.
\item Create \gtr\ opportunities: Insert into $U$ the pair $(t, n\rightarrow n_{rep})$ for each $t\in E$ whose root is $n$, such that the representative of $n$ is $n_{rep}\neq n$.
\end{enumerate}
 \item While ($U$ is not empty)
 \begin{enumerate}
 \item Pop out of $U$ the highest-priority pair $(t,e)$.
\item Apply the corresponding \grow\ or \gtr, resulting in a new tree $t''$;  process($t''$).
\item If $t''$ was not already in $E$, agressively \mergecur:
\begin{enumerate}
\item Let $NT$ be a set of new trees obtained from the \mergecur\ (initially $\emptyset$).
\item Let $\mathbf{p_1}$ be the keyword set of $t''$
\item For each  keyword subset $\mathbf{p_2}$ that is a
  key within $K$, and such that $\mathbf{p_1}\,\cap\, \mathbf{p_2} =\emptyset$
\begin{enumerate}
\item For each tree $t^i$ that corresponds to $\mathbf{p_2}$, try to merge $t''$ with $t^i$. Process any possible result; if it is new (not in $E$ previously), add it to $NT$. \label{item:merge-candidates}
\end{enumerate}
\end{enumerate}
\item Re-plenish $U$ (add more entries in it). This is performed as in step~\ref{item:init-Q} but based on $NT$ (not on $E$). \label{item:replenish-Q}
\end{enumerate}
\end{enumerate}
\end{minipage}}
\caption{Outline of GAM algorithm\label{fig:algo}}
\end{figure*}

We now have the basic exploration steps we need: \grow, \gtr\ and \mergecur.
In this section, we explain how we use them in our integrated keyword search algorithm.

We decide to apply in sequence: one \grow\ or \gtr\ (see below), leading to a new tree $t$, immediately followed by all the \mergecur\ operations possible on $t$. Thus, we call our algorithm \textbf{Grow and Aggressive Merge} (GAM, in short). We merge aggressively in order to detect as quickly as possible when some of our trees, merged at the root, form an answer.

Given that every node of a currently explored answer tree can be connected with several edges, we need to decide which \grow\ (or \gtr) to apply at a certain point. For that, we use a  \textbf{priority queue} $U$ in which we add (tree, edge) entries: for \grow, with the notation above, we add the $(t, e)$ pair, while for \gtr, we add $t$ together with the equivalence edge leading to the representative of $t$'s root. In both cases, when a $(t, e)$ pair is extracted from $U$, we just extend $t$ with the edge $e$ (adjacent to its root), leading to a new tree $t_G$, whose root is the other end of the edge $e$. Then we aggressively merge $t_G$ with all compatible trees explored so far, finally we read from the graph the (data, similarity or equivalence) edges adjacent to $t_G$'s root and add to $U$ more (tree, edge) pairs to be considered further during the search. The algorithm then picks the highest-priority pair in $U$ and reiterates; it stops when $U$ is empty, at a timeout, or when a maximum number of answers are found (whichever comes first).

The last parameter impacting the exploration order is the priority used in $U$: at any point, $U$ gives the highest-priority $(t, e)$ pair, which determines the operations performed next.
\begin{enumerate}
\item Trees matching \emph{many query keywords} are preferable, to go toward complete query answers;
\item At the same number of matched keywords, \emph{smaller trees} are preferable in order not to miss small answers;
\item Finally, among $(t_1,e_1)$, $(t_2,e_2)$ with the same number of nodes and matched keywords,  we prefer the pair with the \emph{higher specificity edge}. 
\end{enumerate}


\noindent\textbf{Algorithm details} Beyond the priority queue $U$ described above, the algorithm also uses a {\em memory of all the trees explored}, called $E$. It also organizes all the (non-answer) trees into a map $K$ in which they can be accessed by the subset of query keywords that they match. The algorithm is shown in pseudocode in Figure~\ref{fig:algo}, following the notations introduced in the above discussion.

While not  shown in Figure~\ref{fig:algo} to avoid clutter, the algorithm {\em only develops minimal trees} (thus, it only finds minimal answers). This is guaranteed:
\begin{itemize}
\item When creating \grow\ and \gtr\ opportunities (steps \ref{item:init-Q} and \ref{item:replenish-Q}): we check not only that the newly added does not close a cycle, but also that the matches present in the new tree satisfy our minimality condition (Section~\ref{sec:search-problem}).
\item Similarly, when selecting potential \mergecur candidates (step \ref{item:merge-candidates}).
\end{itemize}

\section{Experimental evaluation}
\label{sec:evaluation}

We implemented our approach in the \textbf{ConnectionLens} prototype, available
online at \textbf{\url{https://gitlab.inria.fr/cedar/connectionlens}}, which we used to experimentally evaluate the performance of our algorithms.
This section presents the results that we obtained by using synthetic graphs, which are similar to the real-world datasets that we have obtained.
First, we describe the hardware and software setup that we used to run our experiments, and then we give our findings for various combinations of amount of keywords and graph sizes.

\subsection{Hardware and software setup}
\label{sec:evaluation:setup}
We conducted our experiments on a server equipped with 2x10-core Intel
Xeon E5-2640 CPUs clocked at 2.40GHz, and 128GB DRAM. 
The graph is constructed following the approach described in~\cite{construction-paper} and we used Postgres 9.6.5 to store and query the graph for nodes, edges and labels.
The search algorithms are implemented in a Java application which communicates with the database over JDBC, whereas it also maintains an in-memory cache.
Every time that the search algorithm needs information about a node, it first looks into the cache, and if the requested information is not there, it is directly retrieved from the database and then stored in the cache.
To avoid any effects of the cache replacement algorithm, in our
experiments we set the cache to be large enough to include all the
information that has been retrieved from the database.

\noindent\textbf{Synthetic datasets} For controlled experiments, we
generated different types of (RDF) graphs.
The first type is a \emph{line graph}, which the simplest model that we can use.
In the line graph, every node is connected with two others, having one edge for each node, except two nodes which are connected with only one.
By using the line graph, we clearly show the performance of \grow~and \mergecur~operations with respect to the size of the graph.
The second type is a \emph{chain graph}, which is the same as the line graph, but instead of one edge connecting every pair of nodes, we have two.
We use this type to show the performance of the algorithm as we double the amount of edges of the line graph and we give more options to the \grow~and the \mergecur~algorithms.
The third type is the \emph{star graph}, where we have several line graphs connected through a strongly connected cluster of nodes with a representative.
We use this type to show the performance of \gtr, by placing the query keywords on different line graphs.
%
The fourth type is a random graph based on the \emph{Barabasi-Albert}
(BA, in short) model~\cite{Barabasi509}, which generates scale-free networks with only a few nodes (referred to as hubs) of the graph having much higher degree than the rest.
The graph in this model is created in a two-staged process.
During the first stage, a network of some nodes is created.
Then, during the second stage, new nodes are inserted in the graph and they are connected to nodes created during the first stage.
At the second stage, we can control how many connections every node will have to the ones created at the first stage.
By setting that every node created at the second stage to be connected with exactly one node created at the first stage, we have observed that we can construct graphs which are similar to the real-world ones, and therefore we tune our model accordingly.

\textbf{Real-world dataset} The real-world dataset that we used is based on data that we have obtained from journalists with whom we collaborate.
Our dataset combines information on French politics, which we obtained by crawling the web pages of a French newspaper, as we explain in the corresponding Section.

For the microbenchmarks, we report the time needed for our system to return the first answer, as well as the time for all answers.
For the macrobenchmarks, we only report the time to return all the answers, as we do not have full control over the graphs and, hence, it is hard to draw meaningful conclusions and explain them relying on the whole graphs.
Finally, we set an upper bound to the overall execution time at 120 seconds, which is applied to all the experiments that we performed.

\subsection{Querying synthetic datasets}
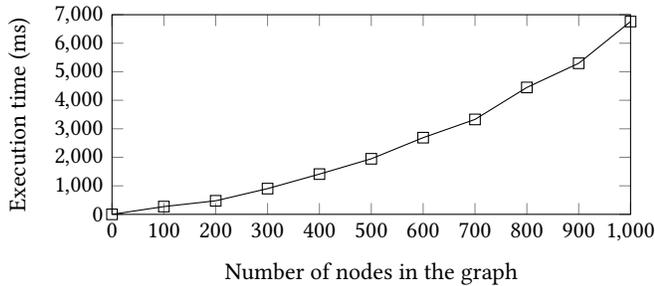
\begin{figure}[t!]
  \centering
  \begin{tikzpicture}
\begin{axis}[
    width=\columnwidth,
    height=0.5\columnwidth,
    xlabel={Number of nodes in the graph},
    ylabel={Execution time (ms)},
    xmin=0, xmax=1000,
    ymin=0, ymax=7000,
    xtick={0,100,200,300,400,500,600,700,800,900,1000},
    ytick={0,1000,2000,3000,4000,5000,6000,7000},
    legend pos=north west,
]

\addplot[
    color=black,
    mark=square,
    ]
    coordinates {
    (0,0)(100,272)(200,475)(300,902)(400,1411)(500,1948)(600,2688)(700,3328)(800,4449)(900,5297)(1000,6760)
    };
\end{axis}
\end{tikzpicture}
\vspace{-2mm}
  \caption{Line graph execution time}
  \label{fig:eval/line}
\vspace{-4mm}
\end{figure}

Figure~\ref{fig:eval/line} shows the execution time of our algorithm when executing a query with two keywords on a line graph, as we vary the number of nodes of the graph.
We place the keywords on the two ``ends'' of the graph to show the impact of the distance on the execution time.
The performance of our algorithm is naturally affected by the size of the graph, as it generates $2*N$ answer trees, where $N$ is the number of nodes.
Given that this is a line graph, there is only one answer, which is the whole graph, and, therefore, the time to find the first answer is also the overall execution time.

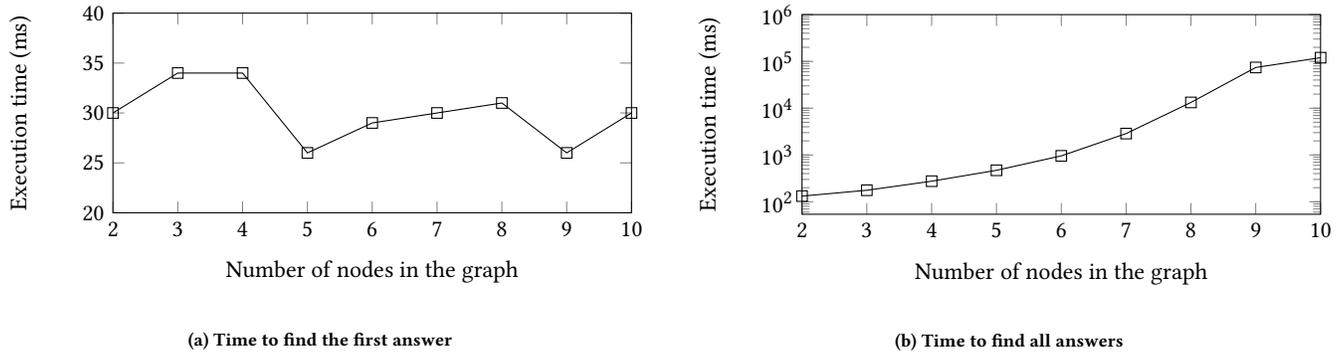
\begin{figure*}[t!]
\subfloat[Time to find the first answer\label{fig:eval/chain/time1}]{
  \begin{minipage}{\columnwidth}
    \begin{tikzpicture}
\begin{axis}[
    width=\textwidth,
    height=0.5\textwidth,
    xlabel={Number of nodes in the graph},
    ylabel={Execution time (ms)},
    xmin=2, xmax=10,
    ymin=20, ymax=40,
    xtick={2,3,4,5,6,7,8,9,10},
    ytick={20,25,30,35,40},
    legend pos=north west,
]

\addplot[
    color=black,
    mark=square,
    ]
    coordinates {
    (2,30)(3,34)(4,34)(5,26)(6,29)(7,30)(8,31)(9,26)(10,30)
    };
\end{axis}
\end{tikzpicture}
  \end{minipage}
}
\hfill
\subfloat[Time to find all answers\label{fig:eval/chain/timeAll}]{
  \begin{minipage}{\columnwidth}
    \begin{tikzpicture}
\begin{axis}[
    width=\textwidth,
    height=0.5\textwidth,
    xlabel={Number of nodes in the graph},
    ylabel={Execution time (ms)},
    xmin=2, xmax=10,
    ymin=0, ymax=1000000,
    xtick={2,3,4,5,6,7,8,9,10},
    ytick={0,10,100,1000,10000,100000,1000000},
    legend pos=north west,
    ymode=log,
]

\addplot[
    color=black,
    mark=square,
    ]
    coordinates {
    (2,132)(3,176)(4,275)(5,470)(6,957)(7,2866)(8,13272)(9,74147)(10,120018)
    };
\end{axis}
\end{tikzpicture}
  \end{minipage}
}
\vspace{-2mm}
\caption{Chain graph execution time}
\vspace{-4mm}
\label{fig:eval/chain}
\end{figure*}

Figure~\ref{fig:eval/chain} shows the execution of our algorithm on a chain graph.
Specifically, Figure~\ref{fig:eval/chain/time1} shows the time elapsed until the first answer is found, whereas Figure~\ref{fig:eval/chain/timeAll} shows the overall execution time.
The execution times reported in Figure~\ref{fig:eval/chain/time1} are almost the same, as the size of the graph increases slowly.
On the other hand, the overall execution times increase at a much
higher (exponential) rate, as shown in
Figure~\ref{fig:eval/chain/timeAll}, where the $y$ axis has a
logarithmic scale. 
The reason is that every pair of nodes is connected with two edges, which increases the amount of answers exponentially with the amount of nodes in the graph.

\begin{figure*}[t!]
\subfloat[Time to find the first answer\label{fig:eval/star/time1}]{
  \begin{minipage}{\columnwidth}
    \begin{tikzpicture}
\begin{axis}[
    width=\textwidth,
    height=0.5\textwidth,
    xlabel={Number of branches in the graph},
    ylabel={Execution time (ms)},
    xmin=2, xmax=14,
    ymin=0, ymax=100000,
    xtick={2,3,4,5,6,7,8,9,10,11,12,13,14},
    ytick={0,10,100,1000,10000,100000},
    legend pos=north west,
    ymode=log,
]

\addplot[
    color=black,
    mark=square,
    ]
    coordinates {
    (2,72)(3,80)(4,82)(5,116)(6,143)(7,197)(8,286)(9,486)(10,964)
    (11,1908)(12,4073)(13,9007)(14,20604)
    };
\end{axis}
\end{tikzpicture}
  \end{minipage}
}
\hfill
\subfloat[Time to find all answers\label{fig:eval/star/timeAll}]{
  \begin{minipage}{\columnwidth}
    \begin{tikzpicture}
\begin{axis}[
    width=\textwidth,
    height=0.5\textwidth,
    xlabel={Number of branches in the graph},
    ylabel={Execution time (ms)},
    xmin=2, xmax=14,
    ymin=0, ymax=1000000,
    xtick={2,3,4,5,6,7,8,9,10,11,12,13,14},
    ytick={0,10,100,1000,10000,100000,1000000},
    legend pos=north west,
    ymode=log,
]

\addplot[
    color=black,
    mark=square,
    ]
    coordinates {
    (2,106)
    (3,131)
    (4,147)
    (5,240)
    (6,400)
    (7,542)
    (8,1168)
    (9,3068)
    (10,9469)
    (11,29628)
    (12,116074)
    (13,119998)
    (14,122771)
    };
\end{axis}
\end{tikzpicture}
  \end{minipage}
}
\vspace{-4mm}
\caption{Star graph execution time}
\vspace{-4mm}
\label{fig:eval/star}
\end{figure*}
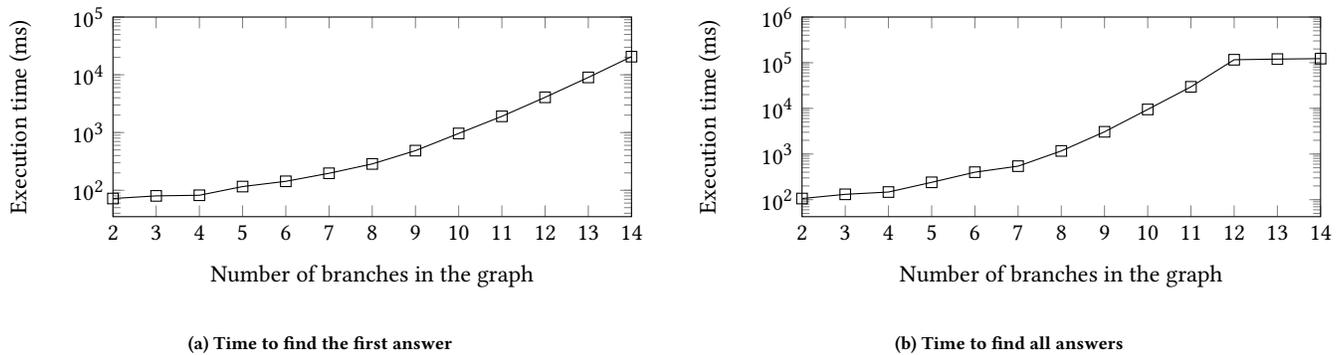

Similar to the chain graph, in Figure~\ref{fig:eval/star} we report the execution time until our algorithm finds the first and all answers (left and right hand side, respectively).
Given that we use keywords which are placed in two different lines connected through the center of the graph, the algorithm has to use \gtr, whereas in the previous cases it only had to use \grow~and \mergecur.
The number of branches, depicted on the $x$ axis of Figure~\ref{fig:eval/star}, corresponds to the number of line graphs connected in the star.
Each line graph has 10 nodes and we place the query keywords at the extremities of two different line graphs.
Given that our algorithm will have to check all possible answers, it follows that the number of merges is exponential to the number of branches, that is $\mathcal{O}(2^K)$, where $K$ is the number of branches.
This behaviour is clearly shown in both parts of
Figure~\ref{fig:eval/star}, where on the $y$ axis (in logarithmic
scale) we show the times to find the first, and, respectively, all
answers. 
Above 12 branches, the timeout of 120 seconds that we have set is hit and, thus, search is terminated, as shown in Figure~\ref{fig:eval/star/timeAll}.

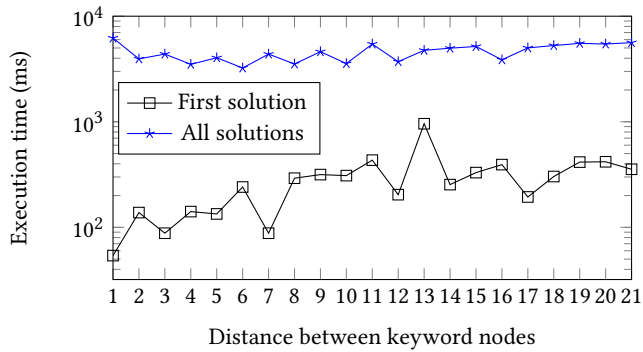
\begin{figure}[ht]
  \centering
  \begin{tikzpicture}
\begin{axis}[
    width=\columnwidth,
    height=0.6\columnwidth,
    xlabel={Distance between keyword nodes},
    ylabel={Execution time (ms)},
    xmin=1, xmax=21,
    ymin=0, ymax=10000,
    xtick={1,2,3,4,5,6,7,8,9,10,11,12,13,14,15,16,17,18,19,20,21},
    ytick={0,10,100,1000,10000},
    legend style={at={(0.4,0.75)}},
    ymode=log,
]

\addplot[
    color=black,
    mark=square,
    ]
    coordinates {
    (1,54)
    (2,138)
    (3,88)
    (4,141)
    (5,134)
    (6,241)
    (7,88)
    (8,292)
    (9,316)
    (10,309)
    (11,433)
    (12,204)
    (13,958)
    (14,254)
    (15,330)
    (16,393)
    (17,194)
    (18,303)
    (19,415)
    (20,418)
    (21,355)
    };

\addplot[
    color=blue,
    mark=star,
    ]
    coordinates {
    (1,6214)
    (2,3943)
    (3,4378)
    (4,3500)
    (5,4042)
    (6,3232)
    (7,4373)
    (8,3527)
    (9,4611)
    (10,3558)
    (11,5448)
    (12,3710)
    (13,4751)
    (14,4977)
    (15,5168)
    (16,3865)
    (17,5009)
    (18,5274)
    (19,5530)
    (20,5448)
    (21,5618)
    };

\legend{First solution, All solutions}

\end{axis}

\end{tikzpicture}
\vspace{-8mm}
  \caption{Barabasi-Albert graph execution time}
\vspace{-4mm}
  \label{fig:eval/ba}
\end{figure}

Figure~\ref{fig:eval/ba} depicts the performance of our algorithm when considering the Barabasi-Albert graph model.
In this experiment, we keep the graph with 2000 nodes fixed and we
vary the position of two keywords, by choosing nodes which have a
distance, as given in the $x$ axis; note the logarithmic $y$ axis. Due to the fact that the graph is
randomly generated within the BA model, we note some irregularity in
the time to the first solution, which however grows at a moderate pace
as the distance between the keyword node grows. The overall relation
between the time to the first solution and the total time confirms
that the search space is very large but that most of the exploration
is not needed, since the first solution is found quite fast. 

\subsection{Querying a real-world dataset}

\begin{table*}[h!]
  \centering
  \begin{tabular}{ |c|c|c|c|c| }
    \hline
    Query keyword(s) & Answers & Answer trees & Time to 1st (ms) & Total time (ms) \\ [0.5ex]
    \hline\hline
    Macron & 118 & 0 & 179 & 390 \\
    \hline
    Trump & 10 & 0 & 26 & 36 \\
    \hline
    Melenchon &8 & 0 & 31 & 39 \\
    \hline
    Christophe, Dettinger & 1105 & 319611 & 136 & 123932 \\
    \hline
    Etienne, Chouard, Rodrigues & 1 & 194 & 144 & 146 \\
    \hline
    Thierry--Paul, Valette, Drouet & 0 & 300813 & N/A & 120001 \\
    \hline
    Melenchon, Aubry & 9 & 284 & 38 & 929 \\
    \hline
    Castaner, flashball & 17 & 1724 & 61 & 545 \\
    \hline
    Drouet, Levavasseur & 18 & 518 & 145 & 309  \\
    \hline
    Dupont--Aignan, Chalencon & 21 & 1850 & 53 & 393 \\
    \hline
    Estrosi, Castaner & 16 & 2203 & 205 & 529 \\
    \hline
    Alexis, Corbiere, Ruffin & 11 & 3782 & 57 & 1022 \\
    \hline
    Macron, Nunez & 13 & 4107 & 1511 & 1561 \\
    \hline
    Hamon, Drouet & 5 & 421 & 71 & 145 \\
    \hline
    Drouet, Ludosky & 27 & 486 & 43 & 145 \\
    \hline
    Salvini, Ludosky & 17 & 1156 & 111 & 375 \\
    \hline
    Salvini, Chouard & 16 & 3205 & 76 & 710 \\
    \hline
    Corbiere, Drouet & 13 & 2341 & 129 & 673 \\
    \hline
    Cauchy, Drouet & 22 & 516 & 96 & 260 \\
    \hline
    Benalla, Nunez & 15 & 1027 & 199 & 347 \\
    \hline
  \end{tabular}
  \caption{Results with real-world dataset}
  \label{tbl:eval/realworld}
\vspace{-5mm}
\end{table*}

This Section includes the results that we obtained by running our algorithm on real-world data.
Our dataset is a corpus of 462 HTML articles (about 6MB) crawled from the French online newspaper Mediapart with the search keywords ``gilets jaunes" (yellow vests, a protest movement in France over the last year).
We built a graph using these articles which consists of 90626 edges and 65868 nodes, out of which 1525 correpond to people, 1240 to locations and 1050 to organizations.
We query the graph using queries of one, two and three different keywords.

We report our findings in Table~\ref{tbl:eval/realworld}.
The results are not given as a basis comparison, rather than as a proof of concept.
Nevertheless, there are several interesting observations to be made.
First, the amount of answers for every query is generally larger than 1.
We allow several results, as the end users (in our case, the investigative journalists) need to see different connections to reach to potentially interesting conclusions.
Second, there are queries where several answers are found, and the execution is interrupted due to the threshold.
We allow the user to set the threshold, based on the results returned every time.
Third, the answers returned to the user are significantly less than the answer trees discovered, showing the impact of minimality as a requirement for returning an answer.

\section{Related work and Conclusions}
\label{sec:related}

Keyword search (KS, in short) is the method of choice for searching in unstructured (typically text) data, and it is also the best search method for novice users, as witnessed by the enormous success of keyword-based search engines. As databases grew larger and more complex, KS has been proposed as a method for searching {\em also} in structured data~\cite{KS-book}, when users are not perfectly familiar with the data, or to get answers enabled by different tuple connection networks. For \textbf{relational} data, in~\cite{DBLP:conf/vldb/HristidisP02} and subsequent works, tuples are represented as nodes, and two tuples are interconnected only through primary key-foreign key pairs. The graphs that result are thus quite uniform, e.g., they consist of ``Company nodes'', ``Employee nodes'' etc. The same model was considered in~\cite{DBLP:conf/icde/OliveiraSM15,DBLP:conf/icde/SayyadianLDG07,DBLP:conf/sigmod/VuOPT08,DBLP:journals/vldb/YanZITY15,DBLP:journals/debu/YuQC10}; \cite{DBLP:conf/icde/SayyadianLDG07} also establishes links based on similarity (or equality) of constants appearing in different relational attributes. 
As explained in Section~\ref{sec:steiner}, our problem is (much) harder since our trees can traverse edges in both directions, and paths can be (much) longer than those based on PK-FK alone.
\cite{DBLP:journals/vldb/YanZITY15} proposes to incorporate user feedback through active learning to improve the quality of answers in a relational data integration setting. We are working to devise such a learning-to-rank approach for our graphs, also. 

KS has also been studied in \textbf{XML documents}~\cite{guo2003xrank,liu2007identifying}. Here, an answer is defined as a subtree of the original document, whose leaves match the query keywords. This problem is much easier than the one we face, since: ($i$)~an XML document is a tree, guaranteeing just one connection between any two nodes; in contrast, there can be any number of such connections in our graphs; ($ii$)~the maximum size of an answer to a $k$-keywords query is $k\cdot h$ where $h$, the height of an XML tree, is almost always quite small, e.g., $20$ is considered ``quite high''; in contrast, with our bi-directional search, the bound is $k\cdot D$ where $D$ is the diameter of our graph - which can be enormously larger. 

Our \grow\ and \mergecur\ steps are borrowed from \cite{dpbf,blinks}, which address KS for \textbf{graphs}, assuming optimal-substructure which does not hold for us, and single-direction edge traversal.  For \textbf{RDF graphs}~\cite{Elbassuoni:2011:KSO:2063576.2063615,DBLP:journals/tkde/LeLKD14} traverse edges in their direction only; moreover, \cite{DBLP:journals/tkde/LeLKD14} also make strong assumptions on the graph, e.g., that all non-leaf nodes have types, and that there are a small number of types (regular graph). In~\cite{error-tolerant}, the authors investigate a different kind of answers to keyword search, the so-called $r$-clique graphs, which they solve with the help of specific indexes. 

Keyword search across \textbf{heterogeneous datasets} has been previously studied in~\cite{Dong:2007,Li:2008:EEK:1376616.1376706}. However, in these works, {\em each answer comes from a single dataset}, that is, they never consider answers spanning over and combining multiple datasets, such as the one shown in Figure~\ref{fig:example-graph}.

In the literature, \textbf{(G)STP} has been addressed under various \textbf{simplifications} that do not hold in our context. For instance: the quality of a solution exponentially decreases with the tree size, thus search can stop when all trees are under a certain threshold~\cite{DBLP:conf/edbt/BonaqueCGM16}; edges are considered in a single direction~\cite{DBLP:journals/debu/YuQC10,Elbassuoni:2011:KSO:2063576.2063615,DBLP:journals/tkde/LeLKD14}; 
 the score function has the suboptimal-structure property~\cite{dpbf,DBLP:conf/sigmod/LiQYM16} etc. These assumptions reduce the computational cost; in contrast, to leave our options open as to the best score function, we worked to build a feasible solution for the general problem we study.  Some works have focused on finding \textbf{bounded (G)STP approximations}, i.e., (G)STP trees solutions whose score is at most $f$ times lower than the optimal one, e.g.,~\cite{soda1998,DBLP:conf/cikm/GubichevN12}. Beyond the differences between our problem and (G)STP, due notably to the fact that our score is much more general (Section~\ref{sec:score}),  non-expert users find it hard to set  $f$.

Beyond the differences we mentioned above, most of which concern our bidirectional search, and the lack of favorable  hypothesis on the score, our work is the first to study querying of graphs originating from {\em integrating several data sources}, while at the same time {\em preserving the identity of each node from the original document}; this is a requirement for integrating, and simultaneously preserving, datasets of journalistic interest. In a companion paper~\cite{construction-paper} we present our latest algorithms for creating such graphs, relying also on information extraction, data matching, and named entity  disambiguation; earlier versions were outlined in~\cite{Chanial2018,cordeiro:hal-02559688}. 


\vspace{1mm}
\noindent\textbf{Acknowledgements}
The authors would like to thank:
Helena Galhardas and Julien Leblay who contributed to previous
versions on this work~\cite{Chanial2018,cordeiro:hal-02559688} and 
Tayeb Merabti for his support in the development and maintenance of
the ConnectionLens system~\cite{construction-paper}. 
This work was partially supported by the H2020 research program under
grant agreement nr. 800192, and by the \href{https://sourcessay.inria.fr}{ANR AI Chair SourcesSay}.

\bibliographystyle{ACM-Reference-Format}
\bibliography{the}


\begin{thebibliography}{33}


\ifx \showCODEN    \undefined \def \showCODEN     #1{\unskip}     \fi
\ifx \showDOI      \undefined \def \showDOI       #1{#1}\fi
\ifx \showISBNx    \undefined \def \showISBNx     #1{\unskip}     \fi
\ifx \showISBNxiii \undefined \def \showISBNxiii  #1{\unskip}     \fi
\ifx \showISSN     \undefined \def \showISSN      #1{\unskip}     \fi
\ifx \showLCCN     \undefined \def \showLCCN      #1{\unskip}     \fi
\ifx \shownote     \undefined \def \shownote      #1{#1}          \fi
\ifx \showarticletitle \undefined \def \showarticletitle #1{#1}   \fi
\ifx \showURL      \undefined \def \showURL       {\relax}        \fi
\providecommand\bibfield[2]{#2}
\providecommand\bibinfo[2]{#2}
\providecommand\natexlab[1]{#1}
\providecommand\showeprint[2][]{arXiv:#2}

\bibitem[\protect\citeauthoryear{Barab{\'a}si and Albert}{Barab{\'a}si and
  Albert}{1999}]%
        {Barabasi509}
\bibfield{author}{\bibinfo{person}{Albert-L{\'a}szl{\'o} Barab{\'a}si} {and}
  \bibinfo{person}{R{\'e}ka Albert}.} \bibinfo{year}{1999}\natexlab{}.
\newblock \showarticletitle{Emergence of Scaling in Random Networks}.
\newblock \bibinfo{journal}{\emph{Science}} \bibinfo{volume}{286},
  \bibinfo{number}{5439} (\bibinfo{year}{1999}).
\newblock
\showISSN{0036-8075}
\urldef\tempurl%
\url{https://doi.org/10.1126/science.286.5439.509}
\showDOI{\tempurl}


\bibitem[\protect\citeauthoryear{Bonaque, Cautis, Goasdou{\'{e}}, and
  Manolescu}{Bonaque et~al\mbox{.}}{2016}]%
        {DBLP:conf/edbt/BonaqueCGM16}
\bibfield{author}{\bibinfo{person}{Rapha{\"{e}}l Bonaque},
  \bibinfo{person}{Bogdan Cautis}, \bibinfo{person}{Fran{\c{c}}ois
  Goasdou{\'{e}}}, {and} \bibinfo{person}{Ioana Manolescu}.}
  \bibinfo{year}{2016}\natexlab{}.
\newblock \showarticletitle{Social, Structured and Semantic Search}. In
  \bibinfo{booktitle}{\emph{EDBT}}.
\newblock


\bibitem[\protect\citeauthoryear{Bugiotti, Bursztyn, Deutsch, Ileana, and
  Manolescu}{Bugiotti et~al\mbox{.}}{2015}]%
        {DBLP:conf/cidr/BugiottiBDIM15}
\bibfield{author}{\bibinfo{person}{Francesca Bugiotti}, \bibinfo{person}{Damian
  Bursztyn}, \bibinfo{person}{Alin Deutsch}, \bibinfo{person}{Ioana Ileana},
  {and} \bibinfo{person}{Ioana Manolescu}.} \bibinfo{year}{2015}\natexlab{}.
\newblock \showarticletitle{Invisible Glue: Scalable Self-Tunning
  Multi-Stores}. In \bibinfo{booktitle}{\emph{CIDR}}.
\newblock


\bibitem[\protect\citeauthoryear{B\v{a}l\v{a}l\v{a}u, Concei\c{a}o, Galhardas,
  Manolescu, Merabti, You, and Youssef}{B\v{a}l\v{a}l\v{a}u
  et~al\mbox{.}}{2020}]%
        {construction-paper}
\bibfield{author}{\bibinfo{person}{Oana B\v{a}l\v{a}l\v{a}u},
  \bibinfo{person}{Catarina Concei\c{a}o}, \bibinfo{person}{Helena Galhardas},
  \bibinfo{person}{Ioana Manolescu}, \bibinfo{person}{Tayeb Merabti},
  \bibinfo{person}{Jingmao You}, {and} \bibinfo{person}{Youssr Youssef}.}
  \bibinfo{year}{2020}\natexlab{}.
\newblock \bibinfo{title}{Graph integration of structured, semistructured and
  unstructured data for data journalism}.
\newblock \bibinfo{howpublished}{Submitted for publication}.
\newblock


\bibitem[\protect\citeauthoryear{Chanial, Dziri, Galhardas, Leblay, Nguyen, and
  Manolescu}{Chanial et~al\mbox{.}}{2018}]%
        {Chanial2018}
\bibfield{author}{\bibinfo{person}{Camille Chanial},
  \bibinfo{person}{R{\'{e}}douane Dziri}, \bibinfo{person}{Helena Galhardas},
  \bibinfo{person}{Julien Leblay}, \bibinfo{person}{Minh~Huong~Le Nguyen},
  {and} \bibinfo{person}{Ioana Manolescu}.} \bibinfo{year}{2018}\natexlab{}.
\newblock \showarticletitle{{\textsc{ConnectionLens}}: Finding Connections
  Across Heterogeneous Data Sources (demonstration)}.
\newblock \bibinfo{journal}{\emph{{VLDB}}} (\bibinfo{year}{2018}).
\newblock


\bibitem[\protect\citeauthoryear{Cheng, Yuan, Li, Chen, and Wang}{Cheng
  et~al\mbox{.}}{2016}]%
        {error-tolerant}
\bibfield{author}{\bibinfo{person}{Yu-Rong Cheng}, \bibinfo{person}{Ye Yuan},
  \bibinfo{person}{Jia-Yu Li}, \bibinfo{person}{Lei Chen}, {and}
  \bibinfo{person}{Guo-Ren Wang}.} \bibinfo{year}{2016}\natexlab{}.
\newblock \showarticletitle{Keyword Query over Error-Tolerant Knowledge Bases}.
\newblock \bibinfo{journal}{\emph{Journal of Computer Science and Technology}}
  \bibinfo{volume}{31} (\bibinfo{year}{2016}).
\newblock
Issue 4.


\bibitem[\protect\citeauthoryear{Cordeiro, Galhardas, Leblay, Manolescu, and
  Merabti}{Cordeiro et~al\mbox{.}}{2020}]%
        {cordeiro:hal-02559688}
\bibfield{author}{\bibinfo{person}{Felipe Cordeiro}, \bibinfo{person}{Helena
  Galhardas}, \bibinfo{person}{Julien Leblay}, \bibinfo{person}{Ioana
  Manolescu}, {and} \bibinfo{person}{Tayeb Merabti}.}
  \bibinfo{year}{2020}\natexlab{}.
\newblock \bibinfo{title}{{Keyword Search in Heterogeneous Data Sources}}.
  (\bibinfo{year}{2020}).
\newblock
\urldef\tempurl%
\url{https://hal.inria.fr/hal-02559688}
\showURL{%
\tempurl}
\newblock
\shownote{Technical report.}


\bibitem[\protect\citeauthoryear{de~Oliveira, da~Silva, and
  de~Moura}{de~Oliveira et~al\mbox{.}}{2015}]%
        {DBLP:conf/icde/OliveiraSM15}
\bibfield{author}{\bibinfo{person}{Pericles de Oliveira},
  \bibinfo{person}{Altigran~Soares da Silva}, {and}
  \bibinfo{person}{Edleno~Silva de Moura}.} \bibinfo{year}{2015}\natexlab{}.
\newblock \showarticletitle{Ranking Candidate Networks of relations to improve
  keyword search over relational databases}. In
  \bibinfo{booktitle}{\emph{IEEE}}.
\newblock


\bibitem[\protect\citeauthoryear{DeWitt, Halverson, Nehme, Shankar,
  Aguilar-Saborit, Avanes, Flasza, and Gramling}{DeWitt et~al\mbox{.}}{2013}]%
        {10.1145/2463676.2463709}
\bibfield{author}{\bibinfo{person}{David~J. DeWitt}, \bibinfo{person}{Alan
  Halverson}, \bibinfo{person}{Rimma Nehme}, \bibinfo{person}{Srinath Shankar},
  \bibinfo{person}{Josep Aguilar-Saborit}, \bibinfo{person}{Artin Avanes},
  \bibinfo{person}{Miro Flasza}, {and} \bibinfo{person}{Jim Gramling}.}
  \bibinfo{year}{2013}\natexlab{}.
\newblock \showarticletitle{Split Query Processing in Polybase}. In
  \bibinfo{booktitle}{\emph{SIGMOD}}.
\newblock
\showISBNx{9781450320375}
\urldef\tempurl%
\url{https://doi.org/10.1145/2463676.2463709}
\showDOI{\tempurl}


\bibitem[\protect\citeauthoryear{Ding, Yu, Wang, Qin, Zhang, and Lin}{Ding
  et~al\mbox{.}}{2007}]%
        {dpbf}
\bibfield{author}{\bibinfo{person}{B. Ding}, \bibinfo{person}{J.~X. Yu},
  \bibinfo{person}{S. Wang}, \bibinfo{person}{L. Qin}, \bibinfo{person}{X.
  Zhang}, {and} \bibinfo{person}{X. Lin}.} \bibinfo{year}{2007}\natexlab{}.
\newblock \showarticletitle{Finding top-$k$ min-cost connected trees in
  databases}. In \bibinfo{booktitle}{\emph{ICDE}}.
\newblock


\bibitem[\protect\citeauthoryear{Dong and Halevy}{Dong and Halevy}{2007}]%
        {Dong:2007}
\bibfield{author}{\bibinfo{person}{Xin Dong} {and} \bibinfo{person}{Alon
  Halevy}.} \bibinfo{year}{2007}\natexlab{}.
\newblock \showarticletitle{Indexing Dataspaces}. In
  \bibinfo{booktitle}{\emph{SIGMOD}}.
\newblock
\showISBNx{978-1-59593-686-8}


\bibitem[\protect\citeauthoryear{Duggan, Elmore, Stonebraker, Balazinska, Howe,
  Kepner, Madden, Maier, Mattson, and Zdonik}{Duggan et~al\mbox{.}}{2015}]%
        {10.1145/2814710.2814713}
\bibfield{author}{\bibinfo{person}{Jennie Duggan}, \bibinfo{person}{Aaron~J.
  Elmore}, \bibinfo{person}{Michael Stonebraker}, \bibinfo{person}{Magda
  Balazinska}, \bibinfo{person}{Bill Howe}, \bibinfo{person}{Jeremy Kepner},
  \bibinfo{person}{Sam Madden}, \bibinfo{person}{David Maier},
  \bibinfo{person}{Tim Mattson}, {and} \bibinfo{person}{Stan Zdonik}.}
  \bibinfo{year}{2015}\natexlab{}.
\newblock \showarticletitle{The BigDAWG Polystore System}.
\newblock \bibinfo{journal}{\emph{SIGMOD Rec.}} \bibinfo{volume}{44},
  \bibinfo{number}{2} (\bibinfo{year}{2015}).
\newblock
\showISSN{0163-5808}
\urldef\tempurl%
\url{https://doi.org/10.1145/2814710.2814713}
\showDOI{\tempurl}


\bibitem[\protect\citeauthoryear{Elbassuoni and Blanco}{Elbassuoni and
  Blanco}{2011}]%
        {Elbassuoni:2011:KSO:2063576.2063615}
\bibfield{author}{\bibinfo{person}{Shady Elbassuoni} {and} \bibinfo{person}{Roi
  Blanco}.} \bibinfo{year}{2011}\natexlab{}.
\newblock \showarticletitle{Keyword Search over RDF Graphs}. In
  \bibinfo{booktitle}{\emph{CIKM}}.
\newblock
\showISBNx{978-1-4503-0717-8}


\bibitem[\protect\citeauthoryear{Garey and Johnson}{Garey and Johnson}{1990}]%
        {garey2011}
\bibfield{author}{\bibinfo{person}{Michael~R. Garey} {and}
  \bibinfo{person}{David~S. Johnson}.} \bibinfo{year}{1990}\natexlab{}.
\newblock \bibinfo{booktitle}{\emph{{Computers and Intractability: A Guide to
  the Theory of NP-Completeness}}}.
\newblock \bibinfo{publisher}{{W. H. Freeman \& Co. New York}}.
\newblock


\bibitem[\protect\citeauthoryear{Garg, Konjevod, and Ravi}{Garg
  et~al\mbox{.}}{1998}]%
        {soda1998}
\bibfield{author}{\bibinfo{person}{N. Garg}, \bibinfo{person}{G. Konjevod},
  {and} \bibinfo{person}{R. Ravi}.} \bibinfo{year}{1998}\natexlab{}.
\newblock \showarticletitle{A polylogarithmic approximation algorithm for the
  group {Steiner} tree problem}. In \bibinfo{booktitle}{\emph{SIAM}}.
\newblock


\bibitem[\protect\citeauthoryear{Gubichev and Neumann}{Gubichev and
  Neumann}{2012}]%
        {DBLP:conf/cikm/GubichevN12}
\bibfield{author}{\bibinfo{person}{Andrey Gubichev} {and}
  \bibinfo{person}{Thomas Neumann}.} \bibinfo{year}{2012}\natexlab{}.
\newblock \showarticletitle{Fast approximation of {Steiner} trees in large
  graphs}. In \bibinfo{booktitle}{\emph{CIKM}}.
\newblock


\bibitem[\protect\citeauthoryear{Guo, Shao, Botev, and Shanmugasundaram}{Guo
  et~al\mbox{.}}{2003}]%
        {guo2003xrank}
\bibfield{author}{\bibinfo{person}{Lin Guo}, \bibinfo{person}{Feng Shao},
  \bibinfo{person}{Chavdar Botev}, {and} \bibinfo{person}{Jayavel
  Shanmugasundaram}.} \bibinfo{year}{2003}\natexlab{}.
\newblock \showarticletitle{{XRANK}: Ranked keyword search over {XML}
  documents}. In \bibinfo{booktitle}{\emph{Proceedings of the 2003 ACM SIGMOD
  international conference on Management of data}}. \bibinfo{pages}{16--27}.
\newblock


\bibitem[\protect\citeauthoryear{He, Wang, Yang, and Yu}{He
  et~al\mbox{.}}{2007}]%
        {blinks}
\bibfield{author}{\bibinfo{person}{Hao He}, \bibinfo{person}{Haixun Wang},
  \bibinfo{person}{Jun Yang}, {and} \bibinfo{person}{Philip~S. Yu}.}
  \bibinfo{year}{2007}\natexlab{}.
\newblock \showarticletitle{{BLINKS:} ranked keyword searches on graphs}. In
  \bibinfo{booktitle}{\emph{SIGMOD}}.
\newblock


\bibitem[\protect\citeauthoryear{Hristidis and Papakonstantinou}{Hristidis and
  Papakonstantinou}{2002}]%
        {DBLP:conf/vldb/HristidisP02}
\bibfield{author}{\bibinfo{person}{Vagelis Hristidis} {and}
  \bibinfo{person}{Yannis Papakonstantinou}.} \bibinfo{year}{2002}\natexlab{}.
\newblock \showarticletitle{{DISCOVER:} Keyword Search in Relational
  Databases}. In \bibinfo{booktitle}{\emph{VLDB}}.
\newblock


\bibitem[\protect\citeauthoryear{Karpathiotakis, Alagiannis, and
  Ailamaki}{Karpathiotakis et~al\mbox{.}}{2016}]%
        {10.14778/2994509.2994516}
\bibfield{author}{\bibinfo{person}{Manos Karpathiotakis},
  \bibinfo{person}{Ioannis Alagiannis}, {and} \bibinfo{person}{Anastasia
  Ailamaki}.} \bibinfo{year}{2016}\natexlab{}.
\newblock \showarticletitle{Fast Queries over Heterogeneous Data through Engine
  Customization}.
\newblock \bibinfo{journal}{\emph{PVLDB}} \bibinfo{volume}{9},
  \bibinfo{number}{12} (\bibinfo{year}{2016}).
\newblock
\showISSN{2150-8097}
\urldef\tempurl%
\url{https://doi.org/10.14778/2994509.2994516}
\showDOI{\tempurl}


\bibitem[\protect\citeauthoryear{Karpathiotakis, Alagiannis, Heinis, Branco,
  and Ailamaki}{Karpathiotakis et~al\mbox{.}}{2015}]%
        {DBLP:conf/cidr/KarpathiotakisA15}
\bibfield{author}{\bibinfo{person}{Manos Karpathiotakis},
  \bibinfo{person}{Ioannis Alagiannis}, \bibinfo{person}{Thomas Heinis},
  \bibinfo{person}{Miguel Branco}, {and} \bibinfo{person}{Anastasia Ailamaki}.}
  \bibinfo{year}{2015}\natexlab{}.
\newblock \showarticletitle{{Just-In-Time Data Virtualization: Lightweight Data
  Management with ViDa}}. In \bibinfo{booktitle}{\emph{CIDR}}.
\newblock


\bibitem[\protect\citeauthoryear{Le, Li, Kementsietsidis, and Duan}{Le
  et~al\mbox{.}}{2014}]%
        {DBLP:journals/tkde/LeLKD14}
\bibfield{author}{\bibinfo{person}{Wangchao Le}, \bibinfo{person}{Feifei Li},
  \bibinfo{person}{Anastasios Kementsietsidis}, {and} \bibinfo{person}{Songyun
  Duan}.} \bibinfo{year}{2014}\natexlab{}.
\newblock \showarticletitle{Scalable Keyword Search on Large {RDF} Data}.
\newblock \bibinfo{journal}{\emph{{IEEE} Trans. Knowl. Data Eng.}}
  \bibinfo{volume}{26}, \bibinfo{number}{11} (\bibinfo{year}{2014}).
\newblock


\bibitem[\protect\citeauthoryear{Li, Ooi, Feng, Wang, and Zhou}{Li
  et~al\mbox{.}}{2008}]%
        {Li:2008:EEK:1376616.1376706}
\bibfield{author}{\bibinfo{person}{Guoliang Li}, \bibinfo{person}{Beng~Chin
  Ooi}, \bibinfo{person}{Jianhua Feng}, \bibinfo{person}{Jianyong Wang}, {and}
  \bibinfo{person}{Lizhu Zhou}.} \bibinfo{year}{2008}\natexlab{}.
\newblock \showarticletitle{{EASE}: An Effective 3-in-1 Keyword Search Method
  for Unstructured, Semi-structured and Structured Data}. In
  \bibinfo{booktitle}{\emph{SIGMOD}}.
\newblock
\showISBNx{978-1-60558-102-6}


\bibitem[\protect\citeauthoryear{Li, Qin, Yu, and Mao}{Li
  et~al\mbox{.}}{2016}]%
        {DBLP:conf/sigmod/LiQYM16}
\bibfield{author}{\bibinfo{person}{Rong{-}Hua Li}, \bibinfo{person}{Lu Qin},
  \bibinfo{person}{Jeffrey~Xu Yu}, {and} \bibinfo{person}{Rui Mao}.}
  \bibinfo{year}{2016}\natexlab{}.
\newblock \showarticletitle{Efficient and Progressive Group {Steiner} Tree
  Search}. In \bibinfo{booktitle}{\emph{SIGMOD}},
  \bibfield{editor}{\bibinfo{person}{Fatma {\"{O}}zcan},
  \bibinfo{person}{Georgia Koutrika}, {and} \bibinfo{person}{Sam Madden}}
  (Eds.).
\newblock


\bibitem[\protect\citeauthoryear{Liu and Chen}{Liu and Chen}{2007}]%
        {liu2007identifying}
\bibfield{author}{\bibinfo{person}{Ziyang Liu} {and} \bibinfo{person}{Yi
  Chen}.} \bibinfo{year}{2007}\natexlab{}.
\newblock \showarticletitle{Identifying meaningful return information for {XML}
  keyword search}. In \bibinfo{booktitle}{\emph{Proceedings of the 2007 ACM
  SIGMOD international conference on Management of data}}.
  \bibinfo{pages}{329--340}.
\newblock


\bibitem[\protect\citeauthoryear{Olston, Reed, Srivastava, Kumar, and
  Tomkins}{Olston et~al\mbox{.}}{2008}]%
        {10.1145/1376616.1376726}
\bibfield{author}{\bibinfo{person}{Christopher Olston},
  \bibinfo{person}{Benjamin Reed}, \bibinfo{person}{Utkarsh Srivastava},
  \bibinfo{person}{Ravi Kumar}, {and} \bibinfo{person}{Andrew Tomkins}.}
  \bibinfo{year}{2008}\natexlab{}.
\newblock \showarticletitle{Pig Latin: A Not-so-Foreign Language for Data
  Processing}. In \bibinfo{booktitle}{\emph{SIGMOD}}
  \emph{(\bibinfo{series}{SIGMOD ’08})}.
\newblock
\showISBNx{9781605581026}
\urldef\tempurl%
\url{https://doi.org/10.1145/1376616.1376726}
\showDOI{\tempurl}


\bibitem[\protect\citeauthoryear{Sayyadian, LeKhac, Doan, and
  Gravano}{Sayyadian et~al\mbox{.}}{2007}]%
        {DBLP:conf/icde/SayyadianLDG07}
\bibfield{author}{\bibinfo{person}{Mayssam Sayyadian}, \bibinfo{person}{Hieu
  LeKhac}, \bibinfo{person}{AnHai Doan}, {and} \bibinfo{person}{Luis Gravano}.}
  \bibinfo{year}{2007}\natexlab{}.
\newblock \showarticletitle{Efficient Keyword Search Across Heterogeneous
  Relational Databases}. In \bibinfo{booktitle}{\emph{ICDE}}.
\newblock


\bibitem[\protect\citeauthoryear{Thusoo, Sarma, Jain, Shao, Chakka, Anthony,
  Liu, Wyckoff, and Murthy}{Thusoo et~al\mbox{.}}{2009}]%
        {10.14778/1687553.1687609}
\bibfield{author}{\bibinfo{person}{Ashish Thusoo}, \bibinfo{person}{Joydeep~Sen
  Sarma}, \bibinfo{person}{Namit Jain}, \bibinfo{person}{Zheng Shao},
  \bibinfo{person}{Prasad Chakka}, \bibinfo{person}{Suresh Anthony},
  \bibinfo{person}{Hao Liu}, \bibinfo{person}{Pete Wyckoff}, {and}
  \bibinfo{person}{Raghotham Murthy}.} \bibinfo{year}{2009}\natexlab{}.
\newblock \showarticletitle{Hive: A Warehousing Solution over a Map-Reduce
  Framework}.
\newblock \bibinfo{journal}{\emph{PVLDB}} \bibinfo{volume}{2},
  \bibinfo{number}{2} (\bibinfo{year}{2009}).
\newblock


\bibitem[\protect\citeauthoryear{Vu, Ooi, Papadias, and Tung}{Vu
  et~al\mbox{.}}{2008}]%
        {DBLP:conf/sigmod/VuOPT08}
\bibfield{author}{\bibinfo{person}{Quang~Hieu Vu}, \bibinfo{person}{Beng~Chin
  Ooi}, \bibinfo{person}{Dimitris Papadias}, {and} \bibinfo{person}{Anthony
  K.~H. Tung}.} \bibinfo{year}{2008}\natexlab{}.
\newblock \showarticletitle{A graph method for keyword-based selection of the
  top-K databases}. In \bibinfo{booktitle}{\emph{SIGMOD}}.
\newblock


\bibitem[\protect\citeauthoryear{Yan, Zheng, Ives, Talukdar, and Yu}{Yan
  et~al\mbox{.}}{2015}]%
        {DBLP:journals/vldb/YanZITY15}
\bibfield{author}{\bibinfo{person}{Zhepeng Yan}, \bibinfo{person}{Nan Zheng},
  \bibinfo{person}{Zachary~G. Ives}, \bibinfo{person}{Partha~Pratim Talukdar},
  {and} \bibinfo{person}{Cong Yu}.} \bibinfo{year}{2015}\natexlab{}.
\newblock \showarticletitle{Active learning in keyword search-based data
  integration}.
\newblock \bibinfo{journal}{\emph{{VLDB} J.}} \bibinfo{volume}{24},
  \bibinfo{number}{5} (\bibinfo{year}{2015}).
\newblock


\bibitem[\protect\citeauthoryear{Yu, Qin, and Chang}{Yu et~al\mbox{.}}{2009}]%
        {KS-book}
\bibfield{author}{\bibinfo{person}{Jeffrey~Xu Yu}, \bibinfo{person}{Lu Qin},
  {and} \bibinfo{person}{Lijun Chang}.} \bibinfo{year}{2009}\natexlab{}.
\newblock \bibinfo{booktitle}{\emph{Keyword Search in Databases}}.
\newblock
\urldef\tempurl%
\url{https://doi.org/10.2200/S00231ED1V01Y200912DTM001}
\showDOI{\tempurl}


\bibitem[\protect\citeauthoryear{Yu, Qin, and Chang}{Yu et~al\mbox{.}}{2010}]%
        {DBLP:journals/debu/YuQC10}
\bibfield{author}{\bibinfo{person}{Jeffrey~Xu Yu}, \bibinfo{person}{Lu Qin},
  {and} \bibinfo{person}{Lijun Chang}.} \bibinfo{year}{2010}\natexlab{}.
\newblock \showarticletitle{Keyword Search in Relational Databases: {A}
  Survey}.
\newblock \bibinfo{journal}{\emph{{IEEE} Data Eng. Bull.}}
  \bibinfo{volume}{33}, \bibinfo{number}{1} (\bibinfo{year}{2010}).
\newblock


\bibitem[\protect\citeauthoryear{Zaharia, Xin, Wendell, Das, Armbrust, Dave,
  Meng, Rosen, Venkataraman, Franklin, Ghodsi, Gonzalez, Shenker, and
  Stoica}{Zaharia et~al\mbox{.}}{2016}]%
        {10.1145/2934664}
\bibfield{author}{\bibinfo{person}{Matei Zaharia}, \bibinfo{person}{Reynold~S.
  Xin}, \bibinfo{person}{Patrick Wendell}, \bibinfo{person}{Tathagata Das},
  \bibinfo{person}{Michael Armbrust}, \bibinfo{person}{Ankur Dave},
  \bibinfo{person}{Xiangrui Meng}, \bibinfo{person}{Josh Rosen},
  \bibinfo{person}{Shivaram Venkataraman}, \bibinfo{person}{Michael~J.
  Franklin}, \bibinfo{person}{Ali Ghodsi}, \bibinfo{person}{Joseph Gonzalez},
  \bibinfo{person}{Scott Shenker}, {and} \bibinfo{person}{Ion Stoica}.}
  \bibinfo{year}{2016}\natexlab{}.
\newblock \showarticletitle{Apache Spark: A Unified Engine for Big Data
  Processing}.
\newblock \bibinfo{journal}{\emph{CACM}} \bibinfo{volume}{59},
  \bibinfo{number}{11} (\bibinfo{year}{2016}).
\newblock
\showISSN{0001-0782}


\end{thebibliography}

\end{document}